\documentclass[a4paper]{amsart}
\usepackage{amscd}
\usepackage{amsxtra} 
\def\opn#1 {\operatorname{#1}}
\def\dopn#1 {
 \def\mYname{\operatorname{#1}}\expandafter\let\csname#1\endcsname=\mYname}

\def\dn#1{\def\arg{#1} \sb{\hbox{$\arg$}}}
\def\br#1{
 \ifx#1<\gdef\Br##1>{\left<##1\right>}\else
 \ifx#1(\gdef\Br##1){\left(##1\right)}\else
 \ifx#1[\gdef\Br##1]{\left[##1\right]}\else
 \ifx#1\{\gdef\Br##1\}{\left\{##1\right\}}\else
 \ifx#1|\gdef\Br##1|{\left|##1\right|}\else
 \ifx#1\|\gdef\Br##1\|{\left\|##1\right\|}\else
 \errmessage{***Bad bracket!}\fi\fi\fi\fi\fi\fi
 \Br}
\let\too=\xrightarrow
\def\seq{\subseteq}
\def\cj{\overline}
\def\ul{\underline}

\long\def\forgetit#1{\relax} 

\newtheorem{thm}{Theorem}[subsection]
\newtheorem{cor}[thm]{Corollary}
\newtheorem{lem}[thm]{Lemma}
\newtheorem{prp}[thm]{Proposition}
\theoremstyle{definition}
\newtheorem{dfn}[thm]{Definition}
\theoremstyle{remark}
\newtheorem{rem}{Remark}    
\newtheorem{rmn}[thm]{Remark} 
\newtheorem{rems}{Remarks}    
\newtheorem{rmns}[thm]{Remarks} 

\long\def\CAR#1#2\NIL{#1}
\long\def\Brm#1\Erm{\edef\nxt{\CAR#1\relax\NIL}\expandafter\ifx\nxt(
 \begin{rems}#1\end{rems}\else
 \begin{rem}#1\end{rem}\fi
}
\long\def\Brmn#1 #2\Erm{
 \edef\nxt{\CAR#2\relax\NIL}
 \expandafter\ifx\nxt(
 \begin{rmns}\label{#1}#2\end{rmns}\else
 \begin{rmn}\label{#1}#2\end{rmn}\fi
}

\numberwithin{equation}{subsection}
\def\nt{\cr} 
\def\Beq#1\Eeq{\begin{equation*} #1 \end{equation*}}
\def\Beqn#1 #2\Eeq{\begin{equation}#2 \label{#1} \end{equation}}
\def\Bml#1\Eml{\begin{multline*} #1 \end{multline*}}
\def\Bmln#1 #2\Eml{\begin{multline}#2 \label{#1} \end{multline}}
\def\Bal#1\Eal{\begin{align*} #1 \end{align*}}
\def\Baln#1 #2\Eal{\begin{align}\label{#1} #2 \end{align}}
\def\Bea#1\Eea{\begin{eqnarray*} #1 \end{eqnarray*}}
\def\Bean#1 #2\Eea{\begin{eqnarray} #2 \label{#1}\end{eqnarray}}
\def\Bcd#1\Ecd{\[\begin{CD} #1 \end{CD}\]}
\def\Bcdn#1 #2\Ecd{
 \begin{equation}\begin{CD}#2 \label{#1}\end{CD}\end{equation}}
\def\bysame{\leavevmode\hbox to3em{\hrulefill}\,} 

\def\even{{\mathbf0}}
\def\odd{{\mathbf1}}
\def\seven{_\even}
\def\sodd{_\odd}
\def\sevR{_{\even,\mathbb R}}
\def\sodR{_{\odd,\mathbb R}}
\def\O{\mathop{\mathcal O}\nolimits}
\def\M{\mathop{\mathcal M}\nolimits}
\dopn L
\def\Cau{^{\opn Cau }}
\def\sol{^{\opn sol }}
\def\rdmm{\mathbb R^{d+1}}
\def\rdm{\mathbb R^d}
\newdimen\mYd  \newbox\mYbox
\def\dcj#1{\def\mYarg{#1}
 \setbox\mYbox=\hbox{$\mYarg$}\mYd=\wd\mYbox
 \vbox{
  \offinterlineskip
  \hbox{\vrule width\mYd height1pt}
  \vskip1pt
  \box\mYbox
}}
\def\mx{_{\opn max }}
\def\G{{\opn G }} 
\def\A{{\opn A }}
\dopn Re
\dopn Im
\dopn CS
\dopn t
\dopn pr
\def\P#1;{{\mathcal P}#1;\ }
\dopn E
\dopn supp
\dopn i
\dopn e
\def\Space{\opn space }
\DeclareMathSymbol{\Subset}       {\mathrel}{AMSa}{"62}
\DeclareMathSymbol{\rightrightarrows}   {\mathrel}{AMSa}{"13}
\def\cfg{^{\opn cfg }}
\def\free{^{\opn free }}
\def\cCau{^{\opn c,Cau }}
\def\csol{^{\opn c,sol }}
\def\cfree{^{\opn c,free }}
\def\CauP{^{\opn Cau' }}

\def\zero{^{\opn zero }}
\def\trp{^{\opn T }} 
\def\exc{^{\opn exc }}
\def\rhs{r.\ h.\ s.}
\def\lhs{l.\ h.\ s.}
\def\mxc{_{\opn max,c }}
\def\mxp{_{\opn max,p }}
\def\ball#1{#1\;\mathbb B}
\def\sd{\tau} 
\def\smloss{\mu} 
\def\cD{{\mathcal D}}
\def\cE{{\mathcal E}}
\def\cEc{{\mathcal E}_{\opn c }}
\def\cEt{{\mathcal E}_{\opn t }}
\def\CauV{^{\opn Cau ,V}}
\def\V{^V}
\def\cH{{\mathcal H}}
\def\bV{{\mathbb V}}
\def\J{{\mathcal J}}
\def\Poinc{\mathfrak P}
\def\cA{{\mathcal A}} 
\def\Psic{\Psi^{\opn c }} 
\def\ulPsic{{\ul\Psi}^{\opn c }}
\def\CData{\phi\Cau}
\def\XiCData{\Xi\sol_{\CData}}

\def\sofe.{s.o.f.e.}
\begin{document}
\def\whatref#1{\cite[#1]{[CMP2]}}  
\def\SFctAnFu{2.5}
\def\CuttingOut{2.12}
\def\FamPhilo{1.11}
\def\StrictSep{Prop. 2.4.4}
\def\CMPref#1{\cite[#1]{[CMP1]}} 
\def\CMPDefAdm{\CMPref{3.1}}     
\def\CMPTransl{\CMPref{3.3}}     
\def\CMPPBil{\CMPref{3.2}}       
\def\CMPFalsIns{\CMPref{3.8}}    
\def\CMPInsMechPrp{\CMPref{Prop. 3.3}} 
\def\CMPDefinesFGerm{\CMPref{Prop. 3.5.2}}

\title[The Cauchy Problem]{
 The Cauchy Problem for Classical Field Equations
\linebreak
 with Ghost and Fermion Fields
}
\author{T. Schmitt}
\thanks{
 Special thanks to the late German Democratic Republic who
 made this research possible by continuous financial support
 over twelve years.
}
\begin{abstract}
Using a supergeometric interpretation of field functionals, we show
that for quite a large class of systems of nonlinear field equations with
anticommuting fields,
infinite-dimensional supermanifolds (smf) of classical solutions can be
constructed. Such systems arise in classical field
models used for realistic quantum field theoretic models.
In particular, we show that under suitable conditions,
the smf of smooth Cauchy data with compact support
is isomorphic with an smf of corresponding classical solutions of the model.
\end{abstract}
\maketitle

{
\tableofcontents
}


\newpage
\section{Systems of field equations}

\subsection{Introduction}
The investigation of the field equations belonging to a
quan\-tum\-field theoretical model as classical nonlinear wave equations
has a long history, dating back to Segal \cite{[Segal]}; cf. also
\cite{[Ginebre/Velo]},
\cite{[Eardley/Moncrief]}, \cite{[ChoYM]}, \cite{[Sniatycki]}. Usually,
Dirac fields have been considered in the obvious way as sections of a
spinor bundle, as e.~g. in \cite{[ChoYM]}.

On the other hand, the rise of supersymmetry made the question of an
adequate treatment of the fermion fields urgent --- supersymmetry and
supergravity do not work with commuting fermion fields. The same applies
to ghost fields: BRST symmetry, which now arouses a considerable interest
among mathematicians (cf. \cite{[KostBRST]}), simply does not exist with
commuting ghost fields.

The anticommutivity required from fermion and ghost fields is often
implemented by letting these fields have their values in the odd part of
an auxiliary Grassmann algebra, as e. g. in \cite{[ChoSugr]}, \cite{[IsBaYa]}.
However, in \cite{[CMP2]}, we have raised our objections against the use
of such an algebra, at least as a fundamental tool. (In \ref{SolValGrass},
we will show how to derive Grassmann-valued solutions from our approach,
which in some sense provides a "universal", intrinsic solution.)

As we have argued in \cite{[CMP2]}, a satisfactory description
of fermion and ghost fields is possible in the framework of
infinite-dimensional supergeometry: the totality of configurations on
space-time should not be considered as a set but as an infinite-dimensional
supermanifold (smf), and the totality of classical solutions should be
a sub-supermanifold. While in \cite{[CMP1]}, \cite{[CMP2]}, we have developed
the necessary supergeometric machinery, this paper will combine it with old
and new techniques in non-linear wave equations in order to implement this
point of view.

In this paper, we consider only two characteristic examples;
a systematic application of our results to a large class of classical
field theories will be given in the successor paper.

Even if the dream of the old, heroic days to construct a quantum field theory
rigorously by direct geometric quantization of the symplectic manifold
of classical solutions (cf. \cite{[Segal]}) has turned out to be
too naive, since it ignores the apparently intrinsic necessity of
renormalization, we nevertheless hope that our construction sheds somewhat
more light onto the geometry of classical field theories. Perhaps, the
dream mentioned will come true some day in a refined variant (cf. also
Remark \ref{MSolFib}).

\subsection{The \protect{$\Phi^4$} toy model}\label{Phi4}\phantom{Some text}
\subsubsection{Classical solutions of Sobolev class}
We start with the usual toy model of every physicist working on
quantum field theory, namely the purely bosonic $\Phi^4$ theory on Minkowski
$\mathbb R^{1+3}$, with the equation of motion
\Beqn 2ndOrdFE
 \bigl({\partial_t}^2 - \sum^3_{a=1}{\partial_a}^2\bigr)\phi
 + m^2\phi +4q\phi^3 =0,
\Eeq
where $m,q\ge0$.
In the usual first-order form, the field equations are
$L_1[\phi_1,\phi_2]=L_2[\phi_1,\phi_2]=0$ where
\Beqn L1L2
 L_1[\Phi_1,\Phi_2] := \partial_t\Phi_1 - \Phi_2, \quad
 L_2[\Phi_1,\Phi_2] :=
 \partial_t\Phi_2 - \sum^3_{a=1}{\partial_a}^2\Phi_1 + m^2\Phi_1 +4q\Phi_1^3.
\Eeq
It is well-known (cf. e. g. \cite[X.13]{[Reed-Simon]}) that for given
Cauchy data $(\phi\Cau_1,\phi\Cau_2)\in H_{k+1}(\mathbb R^3)\oplus H_k(\mathbb
R^3)$
(here $H_k$ is the standard Sobolev space with order $k>1$, in order to ensure
the algebra property of $H_{k+1}$ under pointwise multiplication),
there exists a unique solution
$\phi=(\phi_1,\phi_2)\in C(\mathbb R,H_{k+1}(\mathbb R^3))\otimes\mathbb R^2
 \seq C(\mathbb R^4)\otimes\mathbb R^2$
of the Cauchy problem
\Beq
 L_1[\phi]=L_2[\phi]=0,\quad
  \phi_1(0)=\phi\Cau_1,\quad
  \phi_2(0)=\phi\Cau_2,
\Eeq
and that the arising nonlinear map
\Beq
 \check\Phi\sol:
 M\Cau_k:= H_{k+1}(\mathbb R^3)\oplus H_k(\mathbb R^3) \to
 C(\mathbb R,H_{k+1}(\mathbb R^3))\otimes\mathbb R^2 =: M\cfg_k,\quad
 (\phi\Cau_1,\phi\Cau_2)\mapsto (\phi_1,\phi_2)
\Eeq
is continuous.

As a special case of the general results of this paper, it will turn out
that this map is in fact real-analytic, and that its image is a submanifold of
the Fr\`echet manifold $C(\mathbb R,H_k(\mathbb R^3))\otimes\mathbb R^2$.
Its Taylor expansion at zero arises as the solution of the
"formal Cauchy problem" to find a formal power series
$\Phi\sol[\Phi\Cau_1,\Phi\Cau_2]\in
 \P_f(H_{k+1}(\mathbb R^3)\oplus H_k(\mathbb R^3);
 C(\mathbb R,H_{k+1}(\mathbb R^3))\otimes\mathbb R^2)
$
(cf. \cite{[CMP1]}) with
\Beq
 L_1[\Phi\sol]=L_2[\Phi\sol]=0,\quad
  \Phi\sol_1(0)=\Phi\Cau_1,\quad
  \Phi\sol_2(0)=\Phi\Cau_2
\Eeq
where $\Phi\Cau_1,\Phi\Cau_2$ are "functional variables". This problem is
readily solved by recursion over the degree; one finds
$\Phi\sol =(\Phi_1\sol,\partial_t\Phi_1\sol)$, with
$\Phi_1\sol =\sum_{m\ge0} \Phi\sol_{1,(2m+1)}$,
\Beqn PhiSol
 \begin{array}{rl}
 \Phi\sol_{1,(1)}(t,y) &= \Phi\free_1(t,y) :=
 \int_{\mathbb R^3} dx
  \left(\partial_t\cA(t,y-x)\Phi\Cau_1(x)
  + \cA(t,y-x)\Phi\Cau_2(x)\right),
 \nt
 \\
 \Phi\sol_{1,(3)}(t,y) &=
 4q \int_{\mathbb R\times\mathbb R^3} dsdx\Phi\free_1(s,x)^3 G(t,s,y-x),
 \nt
 \\
 \Phi\sol_{1,(5)}(t,y) &=
  48q^2 \int dsdx\Phi\free_1(s,x)^2 G(t,s,y-x)
   \int ds'dx'\Phi\free_1(s',x')^3 G(s,s',x-x'),
 \nt
 \\
 \Phi\sol_{1,(7)}(t,y) &=
 \int dsdxG(t,s,y-x)
  \Bigl( 192 q^3\Phi\free_1(s,x)
   \left(\int ds"dx"\Phi\free_1(s",x")^3 G(s,s",x-x")\right)^2
   \nt
   \\
\multicolumn{2}{r}{
   \quad +\ 576 q^3\Phi\free_1(s,x)^2 \int ds'dx' \Phi\free_1(x')^2
G(s,s',x-x')
    \int ds"dx"\Phi\free_1(s",x")^3 G(s',s",x'-x")
   \Bigr)
}
  \\
\end{array}
\Eeq
etc. Here $\cA(t,x)$ is the Pauli-Jordan exchange function given by its
spatial Fourier transform as
\Beqn PJFunc
 \hat\cA(t,p) =
 \frac {\sin (\sqrt{m^2 + p^2}t)}{(2\pi)^{3/2}\sqrt{m^2 + p^2}},
\Eeq
and the Green function $G(t,s,x)$ is chosen such that its Cauchy data vanish;
explicitly,
\Beq
 G(t,s,x) = (\theta(t-s) - \theta(-s))\cA(s,x)
\Eeq
where $\theta(\cdot)$  is the Heavyside step function.

The terms of $\Phi\sol_1$ correspond to certain Feynman-like tree diagrams;
for instance, $\Phi\sol_{1,(5)}$ corresponds to the diagram
\par\vskip1cm\par
\unitlength=1.00mm
\linethickness{0.4pt}
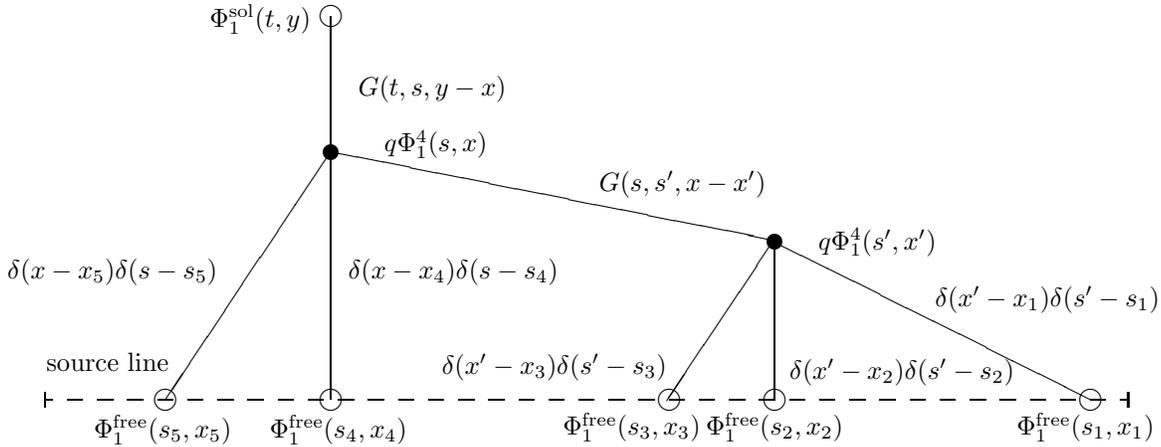
\begin{picture}(144.00,57.41)
\put(110.67,26.00){\makebox(0,0)[cc]{$q\Phi_1^4(s',x')$}}
\put(52.00,39.00){\makebox(0,0)[cc]{$q\Phi_1^4(s,x)$}}
\put(35.34,55.67){\makebox(0,0)[rc]{$\Phi_1\sol(t,y)$}}
\put(41.67,46.33){\makebox(0,0)[lc]{$G(t,s,y-x)$}}
\put(73.67,33.67){\makebox(0,0)[lc]{$G(s,s',x-x')$}}
\put(138.34,1.33){\makebox(0,0)[cc]{$\Phi_1\free(s_1,x_1)$}}
\put(97.00,1.33){\makebox(0,0)[cc]{$\Phi_1\free(s_2,x_2)$}}
\put(78.00,1.33){\makebox(0,0)[cc]{$\Phi_1\free(s_3,x_3)$}}
\put(39.00,1.00){\makebox(0,0)[cc]{$\Phi_1\free(s_4,x_4)$}}
\put(15.67,1.00){\makebox(0,0)[cc]{$\Phi_1\free(s_5,x_5)$}}
\put(8.33,10.00){\makebox(0,0)[cc]{source line}}
\put(118.34,18.33){\makebox(0,0)[lc]{$\delta(x'-x_1)\delta(s'-s_1)$}}
\put(99.00,8.66){\makebox(0,0)[lc]{$\delta(x'-x_2)\delta(s'-s_2)$}}
\put(82.67,9.33){\makebox(0,0)[rc]{$\delta(x'-x_3)\delta(s'-s_3)$}}
\put(40.33,22.00){\makebox(0,0)[lc]{$\delta(x-x_4)\delta(s-s_4)$}}
\put(9.00,22.00){\makebox(0,0)[cc]{$\delta(x-x_5)\delta(s-s_5)$}}
\put(0.00,5.00){\dashbox{2.00}(144.00,0.00)[cc]{}}
\put(16.00,5.00){\circle{2.83}}
\put(38.00,5.00){\circle{2.83}}
\put(83.00,5.00){\circle{2.83}}
\put(97.00,5.00){\circle{2.83}}
\put(38.00,56.00){\circle{2.83}}
\put(38.00,38.00){\circle*{2.00}}
\put(97.00,26.00){\circle*{2.00}}
\put(38.00,56.00){\line(0,-1){18.00}}
\put(38.00,38.00){\line(-2,-3){22.00}}
\put(38.00,38.00){\line(5,-1){59.00}}
\put(97.00,26.33){\line(-2,-3){14.33}}
\put(97.00,26.00){\line(0,-1){21.00}}
\put(97.00,26.00){\line(2,-1){42.00}}
\put(139.00,5.00){\circle{2.83}}
\put(38.00,38.00){\line(0,-1){33.00}}
\end{picture}
\par\vskip1cm\par

The general results of this paper (cf. Thm. \ref{ShortAnal},
Thm. \ref{XiSolXl}, Thm. \ref{MainThm}) now imply:

\begin{cor}
(i) For all $c>0$ there exists $\theta_c>0$ such that $\Phi\sol$,
viewed as power series with values in the Banach space
$C([-\theta_c,\theta_c],\allowbreak H_{k+1}(\mathbb R^3))\otimes\mathbb R^2$,
converges on the $c$-fold unit ball of $M\Cau_k$.

(ii) Fixing Cauchy data $\phi\Cau=(\phi\Cau_1,\phi\Cau_2)\in M\Cau_k$ and
a lifetime $\theta>0$, there exists a neighbourhood $U$ of zero in $M_k\Cau$
such that the translation $\Phi\sol[\Phi\Cau + \phi\Cau]$
(which is only defined for a sufficiently short target time) "prolongates"
to a uniquely determined power series
$\Phi\sol_{\phi\Cau}[\Phi\Cau]$ which
converges on $U$ and solves the field equations.

(iii) The image of the map $\check\Phi\sol: \phi\Cau\mapsto\Phi\sol[\phi\Cau]$
is a submanifold $M\sol_k\seq M_k\cfg$. Moreover, the map
\Beq
 \alpha: M_k\cfg\to M_k\cfg,\quad
 \phi\mapsto (\alpha_1(\phi),\partial_t\alpha_1(\phi)),
 \quad
 \alpha_1(\phi):= \phi_1+\Phi\sol_1[\phi(0)]-\Phi\free_1[\phi(0)]
\Eeq
is an automorphism of $M_k$ which satisfies $\alpha\circ\Phi\free=\Phi\sol$.
\qed\end{cor}

Of course, $\Phi\sol_{\phi\Cau}[\Phi\Cau]$ is just the Taylor expansion of
the map $\Phi\sol$ at $\phi\Cau$.
Note that $U$ may shrink with growing $\theta$; this is connected with the
fact that the target $M_k$ of the map $\check\Phi\sol$ is only a
Fr\`echet space. This indicates that in Minkowski models, there is no way to
work entirely in the framework of Banach spaces.

\subsubsection{Critique and improvement}

Now, viewing $M\sol_k$ as "the" manifold of classical solutions of our model
has the severe defect that we do not know whether it is Lorentz invariant in
a reasonable sense; probably, it is not.

An obvious way out is to use smooth Cauchy data and configurations.
Thm. \ref{MainThmSm} now yields:

\begin{cor}
The restriction of $\check\Phi\sol$ to shooth configuration extends to a
real-analytic map
\Beq
 \check\Phi\sol:
 M_{C^\infty}\Cau:= C^\infty(\mathbb R^3)\otimes\mathbb R^2\to
 C^\infty(\mathbb R^4)\otimes\mathbb R^2=: M\cfg_{C^\infty}.
\Eeq
Its image $M\sol_{C^\infty}$, which is precisely the set of all
smooth solutions of the field equations,
is a submanifold of the Fr\`echet manifold $M\cfg_{C^\infty}$.
\qed\end{cor}

Since the reduction to a first-order system is not Lorentz-invariant,
$M\cfg_{C^\infty}$ is not the adequate configuration space for the purposes
of quantum field theory. One should use instead of it the {\em covariant
configuration space}
\Beq
 M_{C^\infty}:=C^\infty(\mathbb R^4)
\Eeq
and compose $\check\Phi\sol$ with the projection
$M\cfg_{C^\infty}\to M_{C^\infty}$ on the first component; we get:

\begin{cor}
$\check\Phi\sol_1$ restricts to a real-analytic map
\Beq
 \check\Phi\sol_1: M_{C^\infty}\Cau\to M_{C^\infty}.
\Eeq
Its image, which is precisely the set of all
smooth solutions of the original second order field equation \eqref{2ndOrdFE},
is a Lorentz-invariant submanifold of the Fr\`echet manifold $M_{C^\infty}$.
\qed\end{cor}

However, while the absence of any growth condition in spatial direction does
not cause trouble in the construction, due to finite propagation speed,
it causes difficulties in the subsequent investigation of
differential-geometric structures on the image $M\sol_{C^\infty}$:
Every continuous seminorm $p\in\CS(C^\infty(\rdmm))$ is compactly
supported, i. e. there exists some $\Omega\Subset\rdmm$ such that
$p(f)=0$ once $f|_\Omega=0$. This simplifies some proofs
(cf. \ref{BigTrick}), but turns into a vice when looking onto the
superfunctions on $M$: For each superfunction $K\in\O(M_{C^\infty})$,
there exists some compact $\Omega\Subset\rdmm$ such that for the coefficient
functions $K_{k|l}$ of the Taylor expansion at the origin we have
$\supp K_{k|l} \seq \prod^{k+l}\Omega$; analogously for superfunctions on
$M_{C^\infty}\Cau$. Roughly spoken,
$K[\Phi|\Psi]$ is influenced only by the "values" of the fields
on the finite region $\Omega$. In particular, the energy at a given
time instant is not a well-defined superfunction;
only the energy in a finite space-time region is so.

What is still worse, the symplectic structure on
$M_{C^\infty}\Cau\cong M_{C^\infty}\sol$ which one
expects (cf. \whatref{1.12.4} and the successor of this paper)
simply does not make sense; only the induced Poisson structure does.

Thus, it seems reasonable to use only smooth Cauchy data with compact support,
i. e. of test function quality:
$M\Cau:= C^\infty_0(\mathbb R^3)\otimes\mathbb R^2$.
The target space
$M\cfg$ of configurations has to be chosen such that the image of
$\check\Phi\sol$
still is the set of all classical solutions in $M$. Simply taking all smooth
functions on $\mathbb R^4$ which are spatially compactly supported would
violate
Lorentz invariance. However, if we additionally suppose that the spatial
support grows only with light speed then everything is OK:
$M\cfg=C^\infty_c(\mathbb R^4)\otimes\mathbb R^2$ where $C^\infty_c(\mathbb
R^{d+1})$
is the space of all $f\in C^\infty_c(\mathbb R^{d+1})$
such that there exists $R>0$ with $f(t,x) = 0$ for all
$(t,x)\in\mathbb R\times\mathbb R^d$ with $\br|x|\ge \br|t| + R$.
Also, the corresponding covariant configuration space is now
$M=C^\infty_c(\mathbb R^4)$. (Note that this is only a strict inductive limes
of Fr\`echet spaces.)

Thm. \ref{MainThm} now yields:

\begin{cor}\label{Phi4Cor}
$\check\Phi\sol_1$ restricts to a real-analytic map
$\check\Phi\sol_1: M\Cau \to M$.
Its image $M\sol$, which is precisely the set of all those smooth solutions
of the original second order field equation, which have at any time spatially
compact support, is a Lorentz-invariant submanifold of the manifold $M$.
\qed\end{cor}

(Of course, $M\sol$ might miss to contain some interesting classical
solutions; but, at any rate, it comes locally arbitrarily close to them.)

\subsection{The program of this paper}

We start with fixing in \ref{Classmod} a class of systems of classical
nonlinear wave equations in Minkowski space $\rdmm$ which is wide enough to
describe the field equations of many usual models, like e.~g. $\Phi^4$,
quantum electrodynamics, Yang-Mills theory with usual
gauge-breaking term, Faddeev-Popov ghosts, and possibly
minimally coupled fermionic matter. The novelty in our equations is the
appearance of anticommuting fields; in describing the system, they simply
appear as anticommuting variables generating a differential power series
algebra. However, it is no longer obvious what a solution of our system
should be. In fact, as argued in \cite{[CMP2]}, there are no longer
"individual" solutions (besides purely bosonic ones, with all fermionic
components put to zero); but it is sensible to look for {\em families} of
solutions parametrized by supermanifolds. In particular, solutions with
values in Grassmann algebras can be reinterpreted as such families
(cf. \ref{SolValGrass}).

We call a system in our class {\em complete} iff the underlying
bosonic equations admit all-time solutions.
In that case, there is a {\em universal solution family} from which every
other solution family arises in a unique way by pullback. We will construct
this universal solution family by generalizing the map $\check\Phi\sol$ of
Cor. \ref{Phi4Cor} to a {\em morphism of supermanifolds}
\Beqn XiSolIntro
 \check\Xi\sol: M\Cau=\{\text{smf of Cauchy data at $t=0$}\}\longrightarrow
 \{\text{smf of configurations on space-time}\}=M\cfg.
\Eeq
For the construction of this morphism, we follow the usual scheme of solving
non-linear evolution equations: First, one shows the existence of short-time
solutions, and then the existence of all-time solutions.

The necessary supergeometric machinery has been provided in
\cite{[CMP1]}, \cite{[CMP2]}.

Turning to the functional spaces needed, a reasonable choice for
the Cauchy data is the test function space $\cEc\Cau:=\cD(\rdm)$;  for the
configurations we take the space $\cEc$ of all those $f\in C^\infty(\rdmm)$
the support of which on every time slice is compact and grows only with light
velocity (cf. \ref{SpaceCic} for details).

Now we associate to a given model a
{\em configuration supermanifold}, or more precisely, the
{\em supermanifold of smooth configurations with causally growing
spatially compact support}, which is the linear smf
modelled over the "naive configuration space",
\Beq
 M\cfg = \L(\cEc\otimes V);
\Eeq
here $V$ is the target space for the fields.
The standard coordinate (cf. \whatref{\SFctAnFu}) of this linear smf
will be denoted by $\Xi$.

Also, we need the {\em supermanifold of compactly supported smooth Cauchy
data} which is the linear smf
\Beq
 M\Cau = \L(\cD(\rdm)\otimes V)
\Eeq
with the standard coordinate being denoted by $\Xi\Cau$.

We will not use the standard methods of operator semigroups in Hilbert space.
Instead of this, our exposition of infinite-dimensional supergeometry given in
\cite{[IS]} and \cite{[CMP1]}
suggests, and makes here in fact necessary,
another, more direct approach: we expand the solution in a formal,
"functional" power series in the Cauchy data, and then we show convergence
on Sobolev spaces for small times.

Thus, we construct a {\em formal solution} $\Xi\sol[\Xi\Cau]$
of the field equations, which is a formal power series (cf. \CMPref{2.3})
in the Cauchy data $\Xi\Cau$ (in fact, its terms
can be interpreted as belonging to certain tree diagrams).

Next we show that $\Xi\sol[\Xi\Cau](t)$ is for small times
$t$ an {\em analytic} power series on an arbitrarily large multiple of
the unit ball of the Sobolev function space $H_k(\rdm)$
for $k>d/2$. That is, for given $c>0$, there exists $t(c)>0$ with
$\Xi\sol[\Xi\Cau](t)\in\P(H_k(\rdm),\br\|\cdot\|/c; H_k(\rdm))$
(cf. \CMPref{3.2}) for $\br|t|=t(c)$.
Loosely said, this has the consequence that there exist
short time solutions of the field equations: Given bosonic Cauchy data
$\CData$ of Sobolev norm $<c$, a corresponding solution with
lifetime $\ge t(c)$ exists and is given by $\Xi\sol[\CData]$.
Cf. Cor. \ref{LocTraj}.

So far for the short-time solution; next we observe that
a classical solution $\phi\in C([0,\theta],H_k(\rdm)\otimes V\seven)$
of the underlying bosonic equations
with $\theta>0$ can serve as "staircase" to prolongate the
formal solution to an analytic solution in a neighbourhood
of the Cauchy data of $\phi$. That is,
$\Xi\sol[\phi(0)+\Xi\Cau](t,\cdot)$
is analytic up to time $\theta$ (and, in fact, some epsilon beyond).

For proceeding, we have to suppose that the system is {\em causal}, i. e.
that the influence functions have their support in the light cone. In that
case, one can ascend from Sobolev spaces to spaces of smooth functions.

For a causal and complete model, the formal solution $\Xi\sol[\Xi\Cau]$ is
the Taylor expansion of a {\em superfunction}
$\Xi\sol\in\O^{\cEc\otimes V}(M\Cau)$ at zero, and this superfunction
determines the smf morphism \eqref{XiSolIntro} wanted.
This morphism identifies $M\Cau$ with a sub-smf $M\sol$ of $M\cfg$
which we call the {\em supermanifold of classical solutions}. The
name is justified by the fact that given a morphism
$\phi:Z\to M\cfg$, i. e. a $Z$-family of
configurations, it factors through $M\sol$ iff we have
$\phi^*(L_i[\Xi]) =0$, i. e. $Z\too\phi M\cfg$ is a $Z$-family of solutions.

As a variant, we also construct the version
$\check\Xi\sol: M_{C^\infty}\Cau\to M\cfg_{C^\infty}$ which arises by admitting
{\em all} smooth configurations and {\em all} smooth Cauchy data.
By the reasons mentioned in the preceding section, this is not the
functional-analytic quality of main interest.

Another variant arises by considering fluctuations around a fixed
bosonic "background" configuration which solves the bosonic field equations;
cf. \ref{LocExc}.

For the construction of the sub-smf $M\sol$ cut out by the field equations,
the most obvious idea would be to form the ideal subsheaf ${\mathcal J}$ of the
structure sheaf $\O_{M\cfg}$ generated by the superfunctions
$L_i[\Xi](x)$, where $x$ varies over space-time.
Of course, the ideal sheaf algebraically generated by these infinitely many
elements is too small, and one should pass to a suitably completed ideal sheaf.
The main difficulty, however, is that even if a reasonable sub-smf $M\sol$
exists, there is no a priori guaranty that it is equal to
the ringed space $(\supp \O/{\mathcal J},\ \O/{\mathcal J})$.
This is due to a typical infinite-dimensional phenomenon: There
is no general "non-linear Hahn-Banach Theorem", even for a complex-analytic
function on an open subset of a closed linear subspace of a Banach space
it may happen that there is not even locally an extension to a
complex-analytic function defined on an open subset of the ambient space.
Therefore, the approach via ideal sheaves should not play the primary role.
Instead of this, the definition given in \whatref{\CuttingOut}
avoids these difficulties: Given an smf $M$ and some family
$A$ of superfunctions on it, the {\em sub-smf $N$ cut out by $A$} is, if
it exists, uniquely characterized by the requirement that
all elements of $A$ restrict on $N$ to zero, and every smf morphisms
$Z\to M$ which pullbacks all elements of $A$ to zero factors through $N$.
Assertion (v) of Thm. \ref{MainThm} implements this point of view.

{\em A posteori}, it turns out that $M\sol$ is a split sub-smf of $M\cfg$,
and thus we could get it as
$(\supp \O/{\mathcal J},\allowbreak \O/\allowbreak {\mathcal J})$; but this
observation does not help in its construction.

Even in the case of a purely bosonic model, where all our supermanifolds
turn into ordinary real-analytic manifolds modelled over locally convex
spaces, two non-trivial assertions follow from our theory:

First, for any smooth Cauchy datum there exists a short-time solution, and
the latter varies real-analytically with the Cauchy data.

Second, if for any compactly carried smooth Cauchy datum, the existence of
an all-time solution of Sobolev class can be guaranteed, it lies
automatically in $\cEc$ (cf. Lemma \ref{SmCSols}), and in that case, the
all-time solution depends real-analytically on the Cauchy data.

On the other hand, in a purely fermionic model, like e. g.
the Gross-Neveu model, the all-time
solution can be guaranteed to exist a priori; however, there are no
non-trivial "individual" solutions, only families of them.

\subsection{Preliminaries and notations}
Let us shortly recall some notions and conventions from
\cite{[CMP1]}, \cite{[CMP2]}. We follow the usual conventions of
$\mathbb Z_2$-graded algebra: All vector spaces will be $\mathbb Z_2$-graded,
$E=E\seven\oplus E\sodd$ (decomposition into {\em even} and {\em odd} part);
for the {\em parity} of an element, we will write $\br|e|=\mathbf i$ for
$e\in E_{\mathbf i}$. In multilinear expressions, parities add up; this fixes
parities for tensor product and linear maps. (Note that space-time, being not
treated as vector space, remains ungraded. On the other hand, "classical"
function spaces, like Sobolev spaces, are treated as purely even.)

{\em First Sign Rule:} Whenever in a complex multilinear expression
two adjacent terms $a,\ b$ are interchanged the sign $(-1)^{\br|a|\br|b|}$
has to be introduced.

In order to get on the classical level a correct model of operator
conjugation in the quantized theory we also have to use the additional
rules of the hermitian calculus developed in \cite{[HERM]}.
That is, the role of real supercommutative algebras is taken over by
{\em hermitian supercommutative algebras}, i.~e.
complex supercommutative algebras $R$ together with
an involutive antilinear map $\cj{\cdot}: R\to R$
({\em hermitian conjugation\/}) such that
\Beq
 \cj{rs}=\cj s\cdot\cj r
\Eeq
for $r,s\in R$ holds. Note that {\em the real elements of a hermitian
algebra do in general not form a subalgebra}, i. e. $R$ is not just the
complexification of a real algebra. More general, all real vector spaces
have to be complexified before its elements may enter multilinear
expressions.

The essential ingredient of the hermitian framework is the

{\em Second Sign Rule:} If conjugation is applied to a bilinear
expression in the terms $a,\ b$ (i.~e. if conjugation is resolved
into termwise conjugation), either $a,\ b$ have to be rearranged
backwards, or the expression acquires the sign factor $(-1)^{\br|a|\br|b|}$.
Multilinear terms have to be treated iteratively.

Turning to supergeometry, a calculus of real-analytic infinite-dimensional
supermanifolds (smf's) has been constructed by the present author in
\cite{[IS],[CMP1]}.
Here we note that it assigns to every real $\mathbb Z_2$-graded locally
convex space ($\mathbb Z_2$-lcs) $E=E\seven\oplus E\sodd$ a {\em linear
supermanifold} $\L(E)$ which is essentially
a ringed space $\L(E)=(E\seven,\O)$ with underlying topological space
$E\seven$ while the structure
sheaf $\O$ might be thought very roughly of as a kind of completion of
${\mathcal A}(\cdot)\otimes \Lambda E^*\sodd$; here ${\mathcal A}(\cdot)$ is
the sheaf
of real-analytic functions on $E\seven$ while $\Lambda E^*\sodd$ is the
exterior
algebra over the dual of $E\sodd$.

The actual definition of the structure sheaf treats even and odd sector
much more on equal footing than the tensor product ansatz above:
Given a second real $\mathbb Z_2$-graded (lcs) $F$, one defines the space
$\P(E;F)$ of {\em $F$-valued power series on $E$} as the set of all formal
sums $u=\sum_{k,l\ge0} u_{(k|l)}$ where
$u_{(k|l)}: \prod^k E\seven \times \prod^l E\sodd \to F\otimes_{\mathbb R}
\mathbb C$
is a jointly continuous, multilinear map which is symmetric
on $E\seven$ and alternating on $E\sodd$. Now one defines the
sheaf $\O^F(\cdot)$ of {\em $F$-valued superfunctions} on $E\seven$: an
element of $\O^F(U)$ where $U\seq E\seven$ is open is a map $f:U\to\P(E;F)$,
$x\mapsto f_x$, which satisfies
a certain "coherence" condition which makes it sensible to interpret
$f_x$ as the Taylor expansion of $f$ at $x$.

Now the structure sheaf of our ringed space $\L(E)$ is simply
$\O(\cdot):=\O^{\mathbb R}(\cdot)$; it is a sheaf of
hermitian supercommutative algebras,
and each $\O^F(\cdot)$ is a module sheaf over $\O(\cdot)$.

Actually, in considering more general smf's than superdomains,
one has to enhance the structure of a ringed space slightly, in
order to avoid "fake  morphisms" (not every morphism of ringed spaces is
a morphism of supermanifolds). What matters here is that the enhancement is
done in such a way that the following holds (cf. \whatref{Thm. 2.8.1}):

\begin{lem} \label{CoordLem}
Given an $\mathbb Z_2$-lcs $F$ and an arbitrary smf $Z$, the set of morphisms
$Z\to\L(F)$ is in natural 1-1-correspondence with the set
\Beq
 \M^F(Z):= \O^F(Z)_{0,\mathbb R}.
\Eeq
(Here the subscript stands for the real, even part.)
The correspondence works as follows: There exists a distinguished element
$x\in\M^F(\L(F))$ called the {\em standard coordinate}, and one assigns to
$\mu:Z\to\L(F)$ the pullback $\hat\mu:=\mu^*(x)$.
\qed\end{lem}

This is the infinite-dimensional version of the fact that if
$F=\mathbb R^{m|n}$ then a morphism $Z\to\L(\mathbb R^{m|n})$ is known by
knowing
the pullbacks of the coordinate superfunctions, and these can be
prescribed arbitrarily as long as parity and reality are OK.

The most straightforward way to do the enhancement mentioned is a chart
approach; since the supermanifolds we are going to use are actually
all superdomains, and only the morphisms between them are non-trivial, we
need not care here for details.

If $E$, $F$ are spaces of generalized functions on $\rdm$ which contain
the test functions as dense subspace then the Schwartz kernel theorem tells
us that the multilinear forms $u_{(k|l)}$ are given by their integral
kernels, which are generalized functions. Thus one can apply
rather suggestive integral writings (cf. \cite{[CMP1]}) quite analogous to that
used in \eqref{PhiSol}: The general form of a
power series in $\rdm$ is
\Bmln GenPowSer
 K[\Phi|\Psi] = \sum_{k,l\ge0} \frac 1{k!l!} \sum_{I,J}
 \int_{\mathbb R^{d(k+l)}} dx_1\cdots dx_kdy_1\cdots dy_l
 \\
 K^{i_1,\dots,i_k|j_1,\dots,j_l}(x_1,\dots,x_k|y_1,\dots,y_l)
 \Phi_{i_1}(x_1)\cdots\Phi_{i_k}(x_k)\Psi_{j_1}(y_1)\cdots\Psi_{j_l}(y_l)
\Eml
where we have used collective indices $i=1,\dots,N\seven$ and
$j=1,\dots,N\sodd$ for the real components of bosonic and fermionic
fields, respectively. The {\em coefficient functions}
$K^{i_1,\dots,i_k|j_1,\dots,j_l}(x_1,\dots,x_k|y_1,\dots,y_l)$
are distributions which can be supposed to be
symmetric in the pairs $(x_1,i_1),\dots,(x_k,i_k)$ and antisymmetric in
$(y_1,j_1),\dots,(y_l,j_l)$. Of course, they have to satisfy also certain
growth and smoothness conditions. However, what matters here is that the
$\Phi$'s and $\Psi$'s can be formally treated as commuting and anticommuting
fields, respectively; in fact, after establishing the proper calculational
framework, the writing \eqref{GenPowSer} is sufficiently correct.
Also, it is possible to substitute power series into each other
under suitable conditions. Cf. \cite{[CMP1]} for a detailed exposition.

We conclude with some additional preliminaries.
It will be convenient to work not with the bidegrees $(k|l)$
of forms but with {\em total degrees}: For any formal power series
$K\in\P_f(E;F)$ set for $m\ge0$
\Beq
 K_{(m)} :=\sum_{k=0}^m K_{(k|m-k)},\qquad
 K_{(\le m)} :=\sum_{n=0}^m K_{(n)}.
\Eeq
Thus $K=\sum_{m\ge0}K_{(m)}$.

Let $E$ be a $\mathbb Z_2$-lcs and $p\in\CS(E)$ be a continuous seminorm on
$E$;
let $U\seq E$ be the unit ball of $p$. Also, suppose that $F$ is a
$\mathbb Z_2$-graded Banach space. We will use often the suggestive notation
\Beq
 \P(E,U;F) := \P(E,p;F)
\Eeq
(cf. \cite{[CMP1]} for the definition of the \rhs)
for the {\em space of power series converging on $U$}. Indeed, every element
$K\in\P(E,p;F)$ is "a function element on $U\cap E\seven$", i.~e. it is
the Taylor expansion at zero of a uniquely determined superfunction
$K\in\O^F(U\cap E\seven)$ within the superdomain $\L(E)$.

As usual, we call a power series (in the finite-dimensional sense)
in even and odd variables,
$P[y|\eta]=\sum P_{\mu\nu}y^\mu\eta^\nu \in
 \mathbb C[[y_1,\dots,y_m|\eta_1,\allowbreak \dots,\allowbreak \eta_n]]$,
{\em entire} iff for all $R>0$ there exists $C>0$  such that
\Beq
  \br|P_{\mu\nu}| \le C R^{-|\mu|}
\Eeq
for all $\mu,\nu$. The following Lemma is elementary.

\begin{lem}\label{PSEst}
Let  $P[y|\eta]$ be an entire power series of lower degree $\ge l\ge0$.
Then, for any $R>0$ there exists some $C_R>0$
such that if $A$ is a real $\mathbb Z_2$-graded commutative Banach algebra, and
\Beq
 y'_1,\dots,y'_m\in A\seven,\quad
 \eta'_1,\dots,\eta'_n\in A\sodd,\quad
 \br\|y'_i\|\le R,\quad \br\|\eta'_i\|\le R
\Eeq
then
\Beq
 \br\|P(y'_1,\dots,y'_m|\eta'_1,\dots,\eta'_n)\| \le C_R\cdot
 \max\{ \br\|y'_1\|^l,\dots,\br\|y'_m\|^l, \br\|\eta'_1\|^l,\dots,
  \br\|\eta'_n\|^l
 \}.
\Eeq
\qed\end{lem}

Since the calculus of differential polynomials of \cite{[CMP1]}
is insufficient to formulate e.~g. the exponential self-interaction
$\exp \ul\Phi$ of the Liouville model, we consider
{\em differential power series} instead. We set
\Beq
 \mathbb C[[\partial^*\ul\Xi]]:= \bigcup_{n>0}
 \mathbb C[[(\partial^\nu\ul\Xi_i
  )_{i=1,\dots,N,\ \nu\in\mathbb Z_+^d,\ \br|\nu|\le n}]]
\Eeq
where, as usual, $\partial^\nu:= \partial_1^{\nu_1}\cdots\partial_d^{\nu_d}$.
As in \cite{[CMP1]}, the underlined letters $\ul\Xi,\ul\Phi,\ul\Psi$ denote
the indeterminates of an algebra of differential polynomials or
differential power series, while the non-underlined letters
$\Xi,\Phi,\Psi$ denote superfunctions or their Taylor expansions.

As usual, we will write
\Beq
  xy:=\sum_{a=1}^d x_ay_a,\quad x^2:= xx,\quad \br|x| := \sqrt{x^2}
\Eeq
for $x,y\in\rdm$.

All Fourier transformations will be with respect to the spatial
coordinates: For $f\in {\mathcal S}(\rdm)$, we set
\Beq
 \hat f(p) = {\mathcal F}_{x\to p}f(p)
 = (2\pi)^{-d/2}\int_{\rdm} dx \e^{-\i px}f(x);
\Eeq
the extension to ${\mathcal F}: {\mathcal S}'(\rdm)\to {\mathcal S}'(\rdm)$
is done as usual.

\subsection{Systems of field equations}\label{Classmod}
In order to fix a system of field equations we need the following data:

\medskip

I. The {\em space dimension} $d\ge0$; the points of space-time
$\rdmm=\mathbb R\times\rdm$ will be labelled $(t,x)=(t,x_1,\dots,x_d)$.

\medskip

II. The numbers $N\seven,\ N\sodd$ of {\em bosonic}, commuting, and
{\em fermionic},  anticommuting, field components, respectively;
write $N:=N\seven+N\sodd$ for
the total number of field components.

Thus, the setup (cf. \CMPref{2.2}) for superfunctionals on the
fields in space-time will be $(d+1,V)$ with the target space
\Beq
 V := \mathbb R^{N\seven|N\sodd}.
\Eeq
For the tuple of real field components, we will write as in \cite{[CMP1]}
\Beq
 \Xi =(\Xi_1,\dots,\Xi_N) = (\Phi_1,\dots,\Phi_{N\seven}|
 \Psi_1,\dots,\Psi_{N\sodd});
\Eeq
this will be also the functional coordinate on the configuration smf $M\cfg$.

Turning to the Cauchy data, the setup for superfunctionals on them
will be $(d,V)$. We will use the functional coordinates
\Beq
 \Xi\Cau = (\Xi\Cau_1,\dots,\Xi\Cau_N) =
 (\Phi\Cau_1,\dots,\Phi\Cau_{N\seven}|\Psi\Cau_1,\dots,\Psi\Cau_{N\sodd})
\Eeq
for the fields at $t=0$.

\medskip

III. The vector $\sd=(\sd_i)_{i=1}^N\in \mathbb Z_+^N$ of
{\em smoothness offsets}; its role will become clear below.

\medskip

IV. The {\em field equations}, which are given as real, even, entire
differential power series of the form
\Beq
 L_i[\ul\Xi] = \partial_t \ul\Xi_i + \sum_{j=1}^N K_{ij}(\partial_x)\ul\Xi_j
 + \Delta_i[\ul\Xi],\qquad
 \Delta_i[\ul\Xi]\in \mathbb C[[\partial^*\ul\Xi]]_{\even,\mathbb R}.
\Eeq
Here $K_{ij}(\partial_x)$ is a real differential operator with constant
coefficients and containing only spatial derivatives, called the
{\em kinetic operator}, and $\Delta_i[\ul\Xi]$ is
a real, entire differential power series of lower degree $\ge 2$ which is even
and odd
for $i=1,\dots,N\seven$ and $i=N\seven+1,\dots,N\seven+N\sodd$, respectively,
called the {\em interaction term}. We now specify our requirements onto these
terms.

The matrix-valued function
\Beqn DefHatA
 \hat A: \mathbb R\times\mathbb C^d\to\mathbb C^{N\times N},\quad
 \hat A^\Psi(t,p) := (2\pi)^{-d/2}\exp( - K(\i p) t)
\Eeq
satisfies the spatially Fourier-transformed and complexified free field
equations,
\Beq
 \frac d{dt} \hat A (t,p) + K(\i p) \hat A = 0,\quad
 \hat A(0,p)= (2\pi)^{-d/2}1_{N\times N}.
\Eeq
(The reason for the notations $\hat A$ will become clear in the next section.)
We require that there exist $t_0>0$,  $C>0$ such that
\Beqn TheSmCEst
 \br\| \hat A_{ij}(t,p) \| \le C(1+\br|p|)^{\sd_i-\sd_j},
\Eeq
for $p\in\rdm$, $t\in[-t_0,t_0]$.

\Brm
Obviously, the estimate \eqref{TheSmCEst} implies hyperbolicity of the
kinetic operators, i. e. for all $p\in\rdm$, the matrix $K(-\i p)$ has only
imaginary eigenvalues.
\Erm

Define the {\em smoothness degree} of a differential power series
$P=P[\ul\Xi]$ by
\Beq
 \sd(P):=
 \min_{k,\nu} \{\sd_k-\br|\nu|: \quad
   \frac\partial{\partial(\partial^\nu\ul\Xi_k)}
   P\ne0\text{\ \ for some $\nu\in{\mathbb Z_+}^n$}\};
\Eeq
thus, $\sd(\partial^\nu\ul\Xi_k)=\sd_k-l$, and
the smoothness degree of $P$ is just the infimum of the smoothness degrees
of the variables which enter it. Of course, if $P$ is constant
we set $\sd(P)=\infty$.

We have to state a {\em smoothness condition}: For all $i=1,\dots,N$,
we require that
\Beqn SmCond
 \sd_i \le \sd(\Delta_i).
\Eeq

\Brmn MSolFib
Thus, we will assume that numerical values for the coupling constants,
as well as for the masses, have been fixed. In view of the
necessity of renormalization, seemingly intrinsic for any quantization
procedure, it might be sensible to allow these "constants" to vary.
Instead of the solution supermanifold to be constructed we will then get a
bundle of solution smfs over the domain $U\seq\mathbb R^N$ of all tuples of
coupling constants and masses for which the system is complete (cf.
\ref{Cmplness} below). Moreover, the total space of this bundle will
carry a Poisson structure which induces on each fibre a symplectic structure;
perhaps, this is the right object to quantize.
\Erm

\begin{dfn}
A {\em system of field equations (\sofe.)}
is a quadruple $(d,N\seven|N\sodd,\allowbreak \sd,\allowbreak (L_i[\ul\Xi]))$
which satisfies the requirements given above.

Given a \sofe., the {\em underlying bosonic \sofe.} is given by
$(d,N\seven|0,(\sd_1,\dots,\sd_{N\seven}),\allowbreak (L_i[\ul\Phi|0]))$, and
the
{\em underlying free \sofe.} by
$(d,N\seven|N\sodd,\sd,\allowbreak (L_i\free[\ul\Xi]))$
with $L_i\free[\ul\Xi]:= (\partial_t + K(\partial_x))\ul\Xi_j$.
\end{dfn}

We will use matrix writing; in particular, we set
$\Delta=(\Delta_1,\dots,\Delta_N)\trp$ and $L=(L_1,\dots,L_N)\trp$.

\Brm
(1) Usually, the smoothness offsets save that smoothness information which
would be otherwise lost in reducing a temporally higher-order system to a
temporally first-order one.

(2) The smoothness condition is rather constraining; it excludes e. g.
the Korteweg-de Vries equation as well as the nonlinear Schr\"odinger
equations. Fortunately, it is satisfied (for a suitable choice of
smoothness offsets) for apparently all wave equations
occurring in quantum-field theoretical models.
\Erm


\subsection{Function spaces}

In order to keep legibility, we need a certain systematics in the notations:
The superscript "${}\Cau$" will qualify a space as space of Cauchy data,
and thus living on the Cauchy hyperplane $\rdm$;
otherwise, it lives either on $\rdmm$, or, if the notation is
qualified with an argument $I$, on $I\times\rdm$. Also, the superscript
$V$ qualifies as being a space of $V$-valued functions.

Our main technical tool will be the standard Sobolev spaces:
For real $k>d/2$, let $\cH_k\Cau:=H_k(\rdm)$ be the space of all
$f\in L_2(\rdm)$ for which $(1+ |p|)^k\hat f(p)$ is square-integrable.
Recall that by the Sobolev Embedding Theorem, $H_k(\rdm)\seq C^{k'}(\rdm)$
where $k'$ is the maximal integer with $k'<k-d/2$. Also,
$\cH_k\Cau$ is a Banach algebra under pointwise multiplication, cf.
\cite{[Hd2IsAlg]}. (In fact, the consideration of Sobolev spaces with
integer index $k$ would be sufficient for our purposes.)
We equip $H_k(\rdm)$ with the norm
\Beq
 \|f\|_{H_k(\rdm)} := c\int dp (1+ |p|)^{2k}\br|\hat f(p)|^2
\Eeq
where the constant $c>0$ is chosen minimal with the property that
$\|\cdot\|_{H_k(\rdm)}$ is submultiplicative.

The corresponding space of vector-valued Cauchy data is
\Beqn HkCauV
 \cH\CauV_k:= \bigoplus_{i=1}^{N\seven} H_{k+\sd_i}(\rdm)
 \oplus \bigoplus_{i=N\seven+1}^{N\seven+N\sodd} \Pi H_{k+\sd_i}(\rdm)
\Eeq
Set
\Beqn DefM
 \smloss:= \max_{i=1,\dots,N} \max \br\{
 1, \sd_i-\sd(\Delta_i),
  \max_{j=1,\dots,N} \br({\sd_i-\sd_j+\opn ord K_{ij}(\partial_x)})\}
\Eeq
where $\opn ord K_{ij}(\partial_x)$ is the order of the differential operator
($=-\infty$ if $K_{ij}=0$).
The number $\smloss$ will bound the loss of spatial smoothness for
each temporal derivative of the solutions (cf. Prop. \ref{TempSmooth}).

For integer $l\ge0$ such that $k-\smloss l>d/2$, set
\Beq
 \cH_k^l(I) := \bigcap_{n=0}^l C^l(I,H_{k-\smloss n}(\rdm))
\Eeq
(the intersection taken within $C(I\times\rdm)$), equipped with
the topology defined by the seminorms
\Beqn Hkl
 \br\|\xi\|_{\cH_k^l([a,b])}:= \sum_{n=0}^l
 \sup_{t\in[a,b]} \frac 1{n!} \br\|\partial_t^n \xi(t)\|_{H_{k-\smloss
n}(\rdm)}
\Eeq
where $a,b\in I$, $a<b$. We will use mainly the special case that $I$ is
a bounded closed interval; then $\cH_k\V(I)$ is a Banach space, and this
observation justifies the notation on the \lhs\ of \eqref{Hkl}. By
straightforward computation, one has:

\begin{lem}\label{HklLemma}
(i) $\cH^l_k(I)$ is a Banach algebra under pointwise multiplication, and
the norm $\br\|\cdot\|_{\cH_k^l([a,b])}$ is submultiplicative.

(ii) We have a continuous embedding $\cH_k^l(I)\seq C^l(I\times\rdm)$.
\qed\end{lem}

For $k-\smloss l>d/2$ as above, set
\Beqn HkVl
 \cH_k^{l,V}(I) := \bigoplus_{i=1}^{N\seven} \cH_{k+\sd_i}^l(I)
 \oplus \bigoplus_{i=N\seven+1}^{N\seven+N\sodd} \Pi \cH_{k+\sd_i}^l(I)
\Eeq
where $a,b\in I$, $a<b$. Again, $\cH_k^{l,V}(I)$ is a Banach space if
$I$ is a bounded closed interval.

Note that all the spaces qualified with a time definition interval $I$
are not admissible in the sense of \CMPDefAdm\  unless $I=\mathbb R$.
This should not bother us since they appear as target of formal power
series, not as their source.

For better orientation, a table of spaces of configurations
and Cauchy data is given in the Appendix.


\subsection{Influence functions and Green functions}
It follows from \eqref{TheSmCEst} that for fixed $t\in\mathbb R$, the
matrix entries $\hat A_{ij}(t,\cdot)$ are the spatial Fourier transforms
of elements $A_{ij}(t,\cdot)\in {\mathcal S}'(\rdm)$; let
$A(t,\cdot)$ be the corresponding matrix.

For $t\in\mathbb R$, let, as usual, $\theta(t)$ be 1, $1/2$, and 0, for $t>0$,
$t=0$, and $t<0$, respectively. Also, define $f:\mathbb R^2\to\{-1,0,1\}$ by
\Beqn DefFTS
 f(t,s) := \theta(s)\theta(t-s) - \theta(-s)\theta(s-t)
 = \theta(t-s) - \theta(-s).
\Eeq
In particular, $f(t,s)=1$ iff $0<s<t$,
$f(t,s)=-1$ iff $t<s<0$, and $f(t,s)$ vanishes unless
$0\le s\le t$ or $t\le s\le 0$. We define the matrix of {\em Green functions}
$G=(G_{ij}) \in \cD'(\mathbb R^2\times\rdm)^{N\times N}$ by
\Beqn ConstrG
 G(t,s,x)= f(t,s) A(t-s,x).
\Eeq

\Brm
The usual retarded and advanced Green functions are given by
\Beq
 D_{\opn ret }(t,x) =\theta(t) A(t,x),\quad
 D_{\opn adv }(t,x) =-\theta(-t) A(t,x);
\Eeq
conversely,
\Beq
 G(t,s,x) = \theta(s)D_{\opn ret }(t-s,x) +
  \theta(-s)D_{\opn adv }(t-s,x).
\Eeq
\Erm

Altogether, we have
\Beqn KOnG
\begin{array}{ll}
 (\partial_t+K(\partial_x))A(t,x) = 0,
 &\quad A(0,x)=\delta(x)\cdot 1_{N\times N},\\
 (\partial_t+K(\partial_x))G(t,s,x) =
  - \delta(t-s)\delta(x)\cdot 1_{N\times N},
 &\quad G(0,s,x) = 0.
\end{array}
\Eeq

\begin{lem}\label{GreensAndProps}
(i) Given $\xi\Cau\in\cH_k\CauV$ and $g\in\cH_k^{0,V}(\mathbb R)$, the
inhomogeneous linear Cauchy problem to find
$\xi\in {\mathcal S}'(\rdmm)\otimes V$  with
\Beq
 (\partial_t+K(\partial_x))\xi+g=0,\quad
 \xi(0) = \xi\Cau
\Eeq
has the unique solution $\xi = \A\xi\Cau + \G g$ where the operators
$\A,\G$  are given by
\Beq
  \A\xi\Cau(t,y):= \int_{\rdm} dx\; A(t,y-x)\xi\Cau(x),\quad
  \G g(t,y):=\int_{\mathbb R\times\rdm} dsdx\; G(t,s,y-x)g(s,x).
\Eeq

(ii) There exists a constant $C_1$ such that we have for any $\theta>0$
\Beq
 \br\|\A\xi\Cau]\|_{\cH_k^{0,V}([-\theta,\theta])}
  \le C_1\br\|\xi\Cau\|_{\cH_k\CauV}.
\Eeq

(iii) There exists a constant $C_2>0$ such that
\Beq
 \br\|\G g\|_{\cH_k^{0,V}([-\theta,\theta])}
 \le C_2\theta \br\|g\|_{\cH^{0,V}_k([-\theta,\theta])}
\Eeq
for all $\theta>0$.

(iv) The assignment $(\xi\Cau,g)\mapsto \A\xi\Cau + \G g$
defines a continuous linear map
\Beq
 \cH_k\CauV\oplus\cH_k^{0,V}(\mathbb R)\to\cH_k^{0,V}(\mathbb R).
\Eeq
\end{lem}

\begin{proof}
(i) is standard.

Ad (ii).
It follows from \eqref{TheSmCEst} that $\A\xi\Cau(t,\cdot)\in \cH_k\CauV$
for all $t$; however, we have to show that
$\mathbb R\to\cH_k\CauV$,\ \ $t\mapsto\A\xi\Cau(t,\cdot)$ is continuous. We
may assume that the Fourier transform of $\xi\Cau$ has compact support;
now observe that for any $R>0$ we have
$\sup_{\br|p|\le R}\br\|\hat A(t,p) - \hat A(t_0,p)\|\to0$ for $t\to t_0$
where $\br\|\cdot\|$ is any matrix norm. The assertion follows.

Ad (iii). We have $\G g(t,y)= \int ds f(t,s) (\A g(s,\cdot))(t-s,y)$
and hence
\Bml
  \br\|\G g(t,\cdot)\|_{\cH_k\CauV} \le
  \int^\theta_{-\theta} ds \br\| (\A g(s,\cdot))(t-s,\cdot)\|_{\cH_k\CauV}
  \\
  \le C_1 \int^\theta_{-\theta} ds \br\|g(s,\cdot)\|_{\cH_k\CauV}
  \le 2 \theta C_1 \br\|g\|_{\cH^{0,V}_k([-\theta,\theta])}
\Eml
which implies the assertion.

(iv) is an obvious corollary.
\end{proof}

\section{Formal solution and solution families}

\subsection{The formal Cauchy problem}

In the following, we consider the Cauchy problem for the field equations
on the formal power series level. That is, we fix some $k>d/2$
and consider the problem to determine a formal power series in the sense
of \CMPref{2.3},
\Beq
 \Xi\sol = \Xi\sol[\Xi\Cau]\in\P_f(\cH_k\CauV;\cH_k^{0,V}(\mathbb R))\sevR
\Eeq
such that
\Beqn XiSolIsSol
 L[\Xi\sol]=0,\quad \Xi\sol[\Xi\Cau](0,\cdot) = \Xi\Cau(\cdot).
\Eeq
We call this the {\em formal Cauchy problem}.
Its solution $\Xi\sol$, which we call the {\em formal solution of the
\sofe.} will be the power series expansion at
the zero configuration of the solution of the "analytic Cauchy problem",
cf. Cor. \ref{ForSolAndFams}.

Splitting \eqref{XiSolIsSol} into total degrees we get for $n\ge0$
\Beqn DegreeM
 (\partial_t+ K(\partial_x))\Xi\sol_{(n)} + \Delta[\Xi\sol_{(<n)}]_{(n)} = 0,
\Eeq
and hence, using Lemma \ref{GreensAndProps}.(i), $\Xi\sol_{(0)}=0$. Moreover,
using Lemma \ref{GreensAndProps}.(i) again, $\Xi\sol_{(1)}$ is the formal
solution of the free \sofe.:
\Beqn ForSolFirst
 \Xi\sol_{(1)} = \Xi\free[\Xi\Cau] := \A\Xi\Cau.
\Eeq
Finally, for $n>1$, the $\Xi_{(n)}$ are recursively determined by the linear
Cauchy problem consisting of \eqref{DegreeM} and the initial conditions
$\Xi\sol_{(n)}(0)=0$. We get:

\begin{thm}
There exists a uniquely determined solution $\Xi\sol=\sum_{n\ge1}\Xi\sol_{(n)}$
to the formal Cauchy problem. The homogeneous components $\Xi\sol_{(n)}$ are
recursively given by \eqref{ForSolFirst} and
\Beqn SolDegreeM
 \Xi\sol_{(n)} = \G\Delta[\Xi\sol_{(<n)}]_{(n)}
\Eeq
for $n\ge2$. Moreover, $\Xi\sol$ satisfies the integral equation
\Beqn FormalIntEq
 \Xi\sol = \Xi\free + \G\Delta[\Xi\sol].
\Eeq
\qed\end{thm}

\Brm
(1) Note that \eqref{SolDegreeM} makes sense due to Lemma
\ref{GreensAndProps}.(ii).

(2) Of course, $\Xi\sol$, considered as element of the bigger space
$\P_f(\cEc\CauV;\cD'(\rdmm)\allowbreak\otimes V)$, is independent of $k$.

(3) Neither the smoothness condition \eqref{SmCond} nor the analyticity of
the $\Delta_i$ have been used up to now.
\Erm

\subsection{Short-time analyticity of the solution}

\begin{thm} \label{ShortAnal}
Let be given \sofe., and fix $k$ with $k>d/2$. Let
$U\subset\cH_k\CauV$ denote the unit ball. For any $c>0$ there exists
$\theta=\theta_c>0$ such that
\Beqn ShortFrm
 \Xi\sol [\Xi\Cau]\in
 \P(\cH_k\CauV,cU;\cH^{0,V}_k([-\theta,\theta])).
\Eeq
\end{thm}

\begin{proof}
We begin with the estimation of the free solution. From
Lemma \ref{GreensAndProps}.(ii) we have:

\begin{cor}\label{EstXiFree}
There exists a constant $C_1$ such that for all $\theta,c>0$,
\Beq
 \br\|\Xi\free[\Xi\Cau]\|\le C_1c
 \quad\text{within
 $\P(\cH_k\CauV, cU;\cH_k^{0,V}([-\theta,\theta]))$}.
\Eeq
\qed\end{cor}

The idea of the proof of the Theorem is the following: For $n\ge0$, we
have from \eqref{FormalIntEq}
\Beqn IndKey
 \Xi\sol_{(\le n+1)} = \Xi\free + \G\Delta[\Xi\sol_{(\le n)}]_{(\le n+1)}.
\Eeq
We will show that for sufficiently small $\theta>0$ we have for all
$n\ge0$ the estimate
\Beqn IndAss
 \br\|\Xi\sol_{(\le n)}\|\le 2C_1c
 \quad\text{within $\P(\cH\CauV_k,cU;\cH^{0,V}_k([-\theta,\theta]))$.}
\Eeq
Passing to the limit $n\to\infty$ we get assertion (i).

We estimate the interaction term. From Lemma \ref{PSEst} one gets:

\begin{lem}\label{LIntEst}
Given $C'>0$, there exists $C">0$ with the following property:
If $E$ is a $\mathbb Z_2$-lcs, $p\in\CS(E)$  and the power series
$\Xi'=(\Phi'|\Psi')\in \P(E,p;\cH_k^{0,V}([-\theta,\theta]))$
satisfies  $\br\|\Xi'\|<C'$, then
\Beq
 \br\|\Delta[\Xi']\|\le C"
 \quad\text{within $\P(E,p;\cH^{0,V}_k([-\theta,\theta]))$.}
\Eeq
\qed\end{lem}

We now prove \eqref{IndAss} by induction on $n$.
In view of \eqref{ForSolFirst}, the start of induction, $n=1$, is
settled by \eqref{EstXiFree}. Now, for $n\ge1$, we find from
\ref{IndKey} that within $\P(\cH\CauV_k,cU;\cH^{0,V}_k([-\theta,\theta]))$
\Beq
 \br\|\Xi\sol_{(\le n+1)}\| \le \br\|\Xi\free\| +
 \br\|\G\left(\Delta[\Xi\sol_{(\le n)}]_{(\le n+1)}\right)\|.
\Eeq
Using \eqref{EstXiFree} and Lemma \ref{GreensAndProps}.(iii), this
becomes
\Beq
 \le C_1c + C_3\theta\br\|\Delta[\Xi\sol_{(\le n)}]_{(\le n+1)}\|;
\Eeq
because of the hypotheses of induction, the Lemma
applies with $C':=2C_1c $, yielding
\Beq
 \br\|\Xi\sol_{(\le n+1)}\| \le C_1c + C_3\theta C",
\Eeq
and the assertion of induction, $\br\|\Xi\sol_{(\le n+1)}\| \le 2C_1c$,
is satisfied for $\theta\le C_1c/(C_3C")$.

The theorem is proved.
\end{proof}

\subsection{Configuration families}

Let $I$ be a connected subset of $\mathbb R$ with non-empty kernel,
and let $Z$ be an arbitrary smf. A {\em configuration family
of quality $\cH^l_k$ parame\-tri\-zed by $Z$ with time definition domain $I$}
(or {\em $Z$-family over $I$}, for short) is an even, real
superfunction $\Xi'$ on $Z$ with values in the locally convex space
$\cH^{l,V}_k(I)$:
\Beq
 \Xi'=(\Phi'|\Psi')\in\M^{\cH^{l,V}_k(I)}(Z)
\Eeq
(we recall that $\M$ denotes the real, even part of the sheaf $\O$).
Thus, $\Xi'$ encodes a $N$-tuple $\Xi'=(\Xi'_1,\dots,\Xi'_N)$ of
superfunctions $\Xi'_i\in\O^{\cH^l_{k+\sd_i}(I)}(Z)_{\mathbb R}$ on $Z$, with
the $\Phi'_i$ and $\Psi'_j$  being even and odd, respectively.

Beginning with \ref{CausFam}, we will also consider smooth families,
i.~e. families with values in the spaces $\cEc$ and $\cE$ to be
defined later on. (On the other hand, one might consider also still more
general families which have values in spaces of generalized functions;
however, it is then not clear how to define families of solutions).

Now, given an smf morphism $\pi:Z'\to Z$ we can assign to every
$Z$-family $\Xi'$ its {\em pullback} $\Xi":=\pi^*(\Xi')$ which
is a $Z'$-family. In fact, the process of passing from $\Xi'$ to
$\Xi"$ means in family language nothing but a {\em change of parametrization}
(cf. \whatref{\FamPhilo}).

Fixing a $Z$-family $\Xi'$, the field strengthes
$\Xi'_i(t,x)=\delta_{(t,x)}\circ\Xi'_i$ for $(t,x)\in\rdmm$ are
scalar superfunctions on $Z$. More generally, we define the
{\em value at $\Xi'$} of any superfunctional $K\in\O^F(M\cfg)$ as
the pullback of $K$ along $\check{\Xi'}$:
\Beq
 K[\Xi']:=\check{\Xi'}^*(K)\in\O^F(Z).
\Eeq
For instance, in case $Z$ is a point, the value $K[\Xi']$ of
an $F$-valued superfunctional $K$ at a $Z$-family $\Xi'$ is an element of
$F_{\mathbb C}$; thus, for a scalar functional $K\in\O(M\cfg)$, it is
simply a complex number (which, however, is zero for all odd
$K$, and, in particular, for the fermionic field strengthes).

If $Z$ is $0|n$-dimensional then the value $K[\Xi']$ of $K\in\O^F(M\cfg)$
at a $Z$-family $\Xi'$ is an element of $F_{\mathbb C}\otimes_{\mathbb
C}\Lambda_n$
where $\Lambda_n=\mathbb C[\zeta_1,\dots,\zeta_n]$ is a finite-dimensional
Grassmann algebra; thus, for a scalar functional $K\in\O(M\cfg)$, it is a
"Grassmann number" $K[\Xi']\in\Lambda_n$.

\subsection{Solution families}

If $\Xi'\in\M^{\cH^{l,V}_k(I)}(Z)$ is a $Z$-family of
configurations then, using Lemma \ref{HklLemma}.(i) and \eqref{SmCond},
$\Delta[\Xi']\in\O^{\cH^{0,V}_k(I)}(Z)$ is
well-defined. It follows that $L[\Xi']\in\O^{\cD'(I\times\rdm)\otimes V}(Z)$
is well-defined, since the
derivatives needed exist at least in the distributional sense.

We call $\Xi'$ a {\em $Z$-family of solutions},
or {\em solution family} for short, if $L[\Xi']=0$. Obviously, every
pullback of a solution family is a solution family.

Of course, the universal family $\Xi$ is not a family of solutions.
However, we will show in Thm. \ref{MainThm} that in the case of a
complete \sofe., the formal solution will define a family of solutions
$\Xi\sol$ of quality $\cEc$ which is universal for this quality, i.~e.
every other solution family of quality $\cEc$ will be a pullback of
$\Xi\sol$.

If $Z=P$ is a point then a $Z$-family of solutions is just an element
$\phi\in\cH_k^{l,V}(I)\seven$ which solves
the field equations of the underlying bosonic \sofe. in the usual sense.
We call $\phi$ also a {\em trajectory}.

The {\em Cauchy data} of a family $\Xi'\in\M^{\cH_k^{l,V}(I)}(Z)$
with $I\owns0$ is the element $\Xi'(0)\in\M^{\cH_k\CauV}(Z)$.

The condition $L[\Xi']=0$ makes also sense if
$\Xi'$ is only a power series $\Xi'\in\P(E;\cH_k^{l,V}(I))$
where $E$ is an arbitrary $\mathbb Z_2$-lcs; if it is satisfied we call
$\Xi'$ a {\em solution power series}.

Of course, the Taylor expansions
$\Xi"_z$ ($z\in Z$) of any solution family $\Xi"\in\M^{\cH_k^{l,V}(I)}(Z)$ are
solution power series. Conversely, if $I$ is compact then
the target $\cH_k^{l,V}(I)$ is a Banach space, and hence for any element
$\Xi'\in\P(E;\cH_k^{l,V}(I))$ there exists some
$p\in\CS(E)$ such that $\Xi'\in\P(E,p;\cH_k^{l,V}(I))$; by \CMPDefinesFGerm,
any solution power series defines a solution family
$\Xi'\in\M^{\cH_k^{l,V}(I)}(U)$
on the open unit ball $U$ within the superdomain $\L(\hat E_p)$. Hence
we can switch rather freely between solution families and solution power
series.

Spatial Sobolev quality implies to a certain degree temporal differentiabilty:

\begin{prp}\label{TempSmooth}
Every $Z$-family $\Xi'$ of solutions of quality $\cH^{0,V}_k$ is in fact of
quality $\cH^{l,V}_k$ for every $l\ge0$ with $k> \smloss l+ d/2$, where
$\smloss$ is given by \eqref{DefM}.
\end{prp}

\begin{proof}
First, we show the corresponding assertion for solution power series.
We apply induction on $l$; for $l=0$, there is nothing to prove.
For the step from $l$ to $l+1$,
it is sufficient to show that $\Xi'\in\P(E; \cH^{l,V}_k(I))$ implies
\Beqn BTAss
 \partial_t^{l+1}\Xi' \in\P(E; \cH^{0,V}_{k-\smloss(l+1)}(I))
\Eeq
for all $i=1,\dots,N$. From the $i$-th field equation we get
\Beq
 \partial_t^{l+1}\Xi'_i = - {\partial_t}^l
  \sum\nolimits_j K_{ij}(\partial_x)\Xi'_j -{\partial_t}^l \Delta_i[\Xi'].
\Eeq
While clearly ${\partial_t}^l K_{ij}(\partial_x)\Xi'_j\in\P(E;
 \cH^0_{k+ \sd_j - \opn ord K_{ij}(\partial_x) - \smloss l}(I))
 \seq \P(E; \cH^0_{k+ \sd_i - \smloss(l+1)}\allowbreak(I))$, we get
from Lemma \ref{HklLemma}.(i) that
$\Delta_i[\Xi'] \in\P(E; \cH^l_{k+\sd(\Delta_i)}(I))$ for all $j$.
The assertion \eqref{BTAss} follows.

Now, given a solution family $\Xi'\in\M^{\cH^{0,V}_k(I)}(Z)$, it follows that
its Taylor expansions $\Xi'_z$ lie in $\M^{\cH^{l,V}_k(I)}(Z)$.
Using \whatref{\StrictSep}, we get the result.
\end{proof}

$Z$-families of Sobolev quality are uniquely determined by their Cauchy data:

\begin{thm}\label{UniqCor}
Let be given a solution family $\Xi'\in\M^{\cH^{0,V}_k(I)}(Z)$
with $k>d/2$.

(i) If $I\owns0$ then the integral equation
\Beqn FamIntEqu
 \Xi' = \Xi\free[\Xi'(0)] + \G\Delta[\Xi']
\Eeq
holds.

(ii) Suppose that $\Xi"\in\M^{\cH^{0,V}_k(I)}(Z)$ is another solution family
such that for some $t_0\in I$ we have $\Xi'(t_0)=\Xi"(t_0)$.
Then $\Xi'=\Xi"$.
\end{thm}

This can be proved with the ideas of the proof of Thm. \ref{CausUniqThm}
below, so the proof will not be repeated here.

It follows that the Taylor expansions of any solution family arises
from the formal solution by inserting the appropiate Cauchy data:

\begin{cor}\label{ForSolAndFams}
A $Z$-family $\Xi'\in\M^{\cH^{l,V}_k(I)}(Z)$ of configurations
with $k>\smloss l+d/2$ over $I\seq\mathbb R$ is a solution family iff
for each $t_0\in I$ there exists some $\epsilon>0$ such that
\Beq
 \Xi'(t) = \Xi\sol[\Xi'(t_0)](t-t_0)
\Eeq
for $t\in I\cap [t_0-\epsilon,t_0+\epsilon]$.
\qed\end{cor}

Conversely, $\Xi\sol$ produces local solution families:

\begin{cor}
Let be given a family of Cauchy data,
i.~e. an element $\Xi\CauP\in\M^{\cH\CauV_k}(Z)$.
For any compact subset $K\seq \Space Z$ there exists a neighbourhood
$U\supset K$ in $\Space Z$, some $\theta>0$ and and a $U$-family
\Beqn Sought
 \Xi'_U\in\M^{\cH^{0,V}_k([-\theta,\theta])}(U)
\Eeq
of solutions such that
$\Xi'_U(0)=\Xi\CauP$.
\end{cor}

\begin{proof}
It suffices to consider the case that $K=\{z\}$ is a point. In that case,
$\Xi'_U$ can be constructed explicitly as follows: Identify a
neighbourhood of $z$ with a superdomain $V\seq\L(E)$ such that $z$ becomes
the origin, and choose $p\in\CS(E)$ such that
$\Xi\CauP_z\in\P(E,p;\cH\CauV_k)\sevR$; let $c$ be the norm of this element.
By Thm. \ref{ShortAnal}, there exists $\theta>0$ such that
$\Xi\sol [\Xi\Cau]\in
 \P(\cH_k\CauV,\br\|\cdot\|/(c+1);\cH^{0,V}_k([-\theta,\theta]))$.
Then
\Beq
 \Xi'_z:=\Xi\sol[\Xi\CauP_z]\in
 \P(E,p;\cH^{0,V}_k([-\theta,\theta]))\sevR,
\Eeq
and, letting $U$ be the unit ball of $p$ in $E\seven$ and using
\CMPDefinesFGerm, this element determines the solution family
\eqref{Sought} wanted.
\end{proof}

In particular, we get that trajectories over short times
always exist, i.~e. that the even, bosonic field equations are short-time
solvable in the usual sense:

\begin{cor} \label{LocTraj}
Given bosonic Cauchy data $\CData\in\cH\CauV_k(\rdm)\seven$
there exists $\theta>0$ and a unique trajectory
$\phi\in \cH^{0,V}_k([-\theta,\theta])\seven$ with
$\phi(0)=\phi\Cau$. It is given by
\Beq
 \phi = \Xi\sol[\phi\Cau|0].
\Eeq
\qed\end{cor}

\Brm
At this stage, we have no information on whether $\phi$ can be extended
to an all-time trajectory. In fact, we have to suppose this later on,
defining in this way the completeness of a \sofe.
\Erm

Obviously, if $\Xi'\in\M^{\cE\V(I)}(Z)$ is a solution family then so is
every temporal translate $\Xi'(\cdot+t_0)\in\M^{\cE\V(I+t_0)}(Z)$ with
$t_0\in\mathbb R$. (A general discussion of symmetry transformations will
be given in the successor paper.) This gives the possibility to "splice"
solution families: From Thm. \ref{UniqCor}.(ii) and Thm. \ref{ShortAnal}
we get

\begin{cor}\label{SpliceLem}
If $\Xi'\in\M^{\cH_k^{l,V}(I)}(Z)$ and $\Xi"\in\M^{\cH_k^{l,V}(I')}(Z)$
are solution families with $0\in I'$ such that $\Xi'(t_0)=\Xi"(0)$
for some $t_0\in I$ then there exists a unique solution family
$\Xi^{\opn splice }\in\M^{\cH_k^{l,V}(I_1)}(Z)$ with $I_1:= I\cup (t_0+I')$
such that $\Xi^{\opn splice }|_I=\Xi'$,\ \
$\Xi^{\opn splice }|_{t_0+I'}=\Xi"(\cdot+t_0)$.
\qed\end{cor}


\subsection{Lifetime intervals}
Fix a $\mathbb Z_2$-lcs $E$ and $p\in\CS(E)$. A priori, the space
$\P(E,p;\cH_k^{l,V}\allowbreak (I))$  is defined only if $I$
is closed and bounded
(since only in that case, the target is a Banach space); we extend the
definition to any connected subset $I\seq\mathbb R$ with non-empty kernel by
\Beq
 \P(E,p;\cH_k^{l,V}(I)):= \lim_{\longleftarrow}
 \P(E,p;\cH_k^{l,V}([a,b]))\seq\P(E;\cH_k^{l,V}(I)).
\Eeq
For shortness, we will write again $\br\|\cdot\|_F$ for the norms in
$\P(E,p;F)$ where $F$ is one of the Banach spaces $\cH_k^l(I)$,\ \
$\cH_k^{l,V}(I)$,
\ \ $\cH_k\CauV$ with $I$ being closed.

The following Lemma is a standard idea in nonlinear wave equations.

\begin{lem}\label{SelfSm}
Let $k>d/2$, let be given a solution power series
$\Xi'\in\P(E,p;\cH_{k+1}^{l,V}\allowbreak(I))\sevR$
where either $I=[0,b)$ with $0<b<\infty$, or
$I=(b,0]$ with $0>b>-\infty$, and suppose
\Beqn SupHyp
 \sup_{t\in I} \br\|\Xi'(t)\|_{\cH_k\CauV} <\infty.
\Eeq
Then we have also
\Beq
 \sup_{t\in I} \br\|\Xi'(t)\|_{\cH_{k+1}\CauV} <\infty.
\Eeq
\end{lem}

\begin{proof}
We treat only the case $I=[0,b)$ with $0<b<\infty$.
It is sufficient to prove that there exists some $\theta\in(0,b)$
such that $\br\|\Xi'\|_{\cH_{k+1}^{0,V}([\theta,t])}$
remains bounded with varying $t\in [\theta,b)$, which is
equivalent with boundedness of
$\br\|\partial_a\Xi'\|_{\cH_k^{0,V}([\theta,t])}$ for all $a=1,\dots,d$.

We use a time-shifted variant of \eqref{FamIntEqu}: Fix $\theta\in(0,b)$.
We have
$\Xi' = \Xi\free[\Xi'(\theta)] + \Theta$
with
\Beq
 \Theta(t):=\G(\Delta[\Xi'](\cdot+\theta))(t-\theta).
\Eeq
Hence
\Beq
 \br\|\partial_a\Xi'\|_{\cH_k^{0,V}([\theta,t])}
 \le \br\|\partial_a\Xi\free[\Xi'(\theta)]
        \|_{\cH_k^{0,V}([\theta,t])}
 + \br\|\partial_a\Theta\|_{\cH_k^{0,V}([\theta,t])}.
\Eeq
Using partial integration,
$\partial_a\Theta(t):=\G(\partial_a\Delta[\Xi'](\cdot+\theta))(t-\theta)$.
Using Lemma \ref{GreensAndProps}.(iii) we get
\Beqn SlfSm1
 \br\|\partial_a\Theta\|_{\cH_k^{0,V}([\theta,t])} \le
 K_1(b-\theta)  \br\|\partial_a\Delta[\Xi']\|_{\cH^{0,V}_k([\theta,t])}
\Eeq
for all $t\in[\theta,b)$ with some $K_1>0$ which depends only on
the underlying free \sofe.
Now
\Beqn PtlAExpd
  \partial_a\Delta_i[\Xi'] = \sum_{i,j,\nu} \partial_a\partial^\nu\Xi'_j
  \cdot \frac\partial{\partial(\partial^\nu\ul\Xi_j)} \Delta_i[\Xi']
\Eeq
where the sum runs over those $i,j=1,\dots,N$ and $\nu\in {\mathbb Z_+}^d$
for which
$\frac\partial{\partial(\partial^\nu\ul\Xi_j)} \Delta_i[\ul\Xi]\not=0$.
Using Lemma \ref{HklLemma}.(i),
\Beqn DDeltEst
  \br\|\partial_a\Delta[\Xi']\|_{\cH^{0,V}_k([\theta,t])}
  \le
  \sum_{i,j,\nu}
  \br\|\partial_a\partial^\nu\Xi'_j\|_{\cH^0_{k+\sd(\Delta_i)}([\theta,t])}
  \cdot
 \br\|\frac\partial{\partial(\partial^\nu\ul\Xi_j)} \Delta[\Xi']\|
 _{\cH^{0,V}_k([\theta,t])}
\Eeq
with the same range of the sum as in \eqref{PtlAExpd}.
Now, for the $(i,j,\nu)$ which enter \eqref{PtlAExpd}, we
have $\sd(\Delta_i)\le\sd_j-\br|\nu|$, and hence
\Beq
 \br\|\partial_a\partial^\nu\Xi'_j\|_{\cH^0_{k+\sd(\Delta_i)}([\theta,t])}
 \le
 \br\|\partial_a\Xi'_j\|_{\cH^0_{k+\sd(\Delta_i)+\br|\nu|}([\theta,t])}
 \le
 \br\|\partial_a\Xi'_j\|_{\cH^0_{k+\sd_j}([\theta,t])}
 \le
 \br\|\partial_a\Xi'\|_{\cH^{0,V}_k([\theta,t])}.
\Eeq
{}From \eqref{SupHyp}
we get that the second factor on the \rhs\ of \eqref{DDeltEst}
is bounded by a constant $K_2$, so that
\Beqn SlfSm2
 \br\|\partial_a\Delta[\Xi']\|_{\cH^{0,V}_k([\theta,t])}
  \le K_2\cdot\br\|\partial_a\Xi'\|_{\cH^{0,V}_k([\theta,t])}.
\Eeq
Putting \eqref{SlfSm1}, \eqref{SlfSm2} together, and
using also Cor. \ref{EstXiFree}, we get an estimate
\Beq
 \br\|\partial_a\Xi'\|_{\cH_k^{0,V}([\theta,t])}
 \le K_3 \|\Xi'(\theta)\|_{\cH_{k+1}\CauV}
 + K_1K_2(b-\theta)\br\|\partial_a\Xi'\|_{\cH^{0,V}_k([\theta,t])}.
\Eeq
Now, fixing $\theta:= b - 1/(2K_1K_2)$, we get
$\br\|\partial_a\Xi'\|_{\cH_k^{0,V}([\theta,t])}
 \le 2K_3 \|\Xi'(\theta)\|_{\cH_{k+1}\CauV}$
for all $t\in [\theta,b) \forgetit]$ and the assertion.
\end{proof}

\begin{prp}\label{Lt4PEcH}
Let be given a $\mathbb Z_2$-lcs $E$, a seminorm $p\in\CS(E)$,
and an element
\Beqn TheLtCauD
 \Xi\CauP\in\P(E,p;\cH\CauV_k)\sevR
\Eeq
where $k>d/2$.

(i) There exists a uniquely determined pair $(I\mx,\Xi'\mx)$ where
$I\mx\seq\mathbb R$ is connected and open with $I\mx\owns 0$, and
\Beqn LtMxSol
 \Xi'\mx\in\P(E,p;\cH_k^{0,V}(I\mx))\sevR
\Eeq
is a solution power series which has $\Xi\CauP$ as Cauchy data,
$\Xi'\mx(0) = \Xi\CauP$, and is maximal with this property: If
$(I",\Xi")$ is another pair where $I"\mx\seq\mathbb R$ is connected and
open with $I"\mx\owns 0$, and $\Xi"\in\P(E,p;\cH_k^{0,V}(I"))$
is a solution power series which has $\Xi\CauP$ as Cauchy data
then $I"\seq I\mx$, and $\Xi'\mx|_{I"} = \Xi"$.

We call $\Xi'\mx$ the {\em maximal solution power series} belonging to the
Cauchy data \eqref{TheLtCauD}.

(ii) If $a:=\inf_{t\in I\mx} t >-\infty$ then, for any $\epsilon>0$
\Beq
 \operatornamewithlimits{lim\ sup}_{t\to a}
 \|\Xi'\mx(t)\|_{\cH_{d/2+\epsilon}\CauV} = \infty;
\Eeq
likewise, if $b:=\sup_{t\in I\mx} t \allowbreak <\infty$
then
\Beq
 \operatornamewithlimits{lim\ sup}_{t\to b}
 \|\Xi'\mx(t)\|_{\cH_{d/2+\epsilon}\CauV} = \infty.
\Eeq
\end{prp}

\begin{proof}
It follows from Thm. \ref{ShortAnal} and  Thm. \ref{UniqCor} that there
exists at least a connected subset with non-empty kernel $I\mx\seq\mathbb R$
with $I\mx\owns 0$ and a maximal element \eqref{LtMxSol};
we have to prove that $I\mx$ is open.
Let $b:=\sup_{t\in I\mx} t$, and assume that $b\in I\mx$. Then
$\Xi'\mx(b)\in\P(E,p;\cH_k\CauV)$
is well-defined; let $c'$ be its norm. By Thm. \ref{ShortAnal}, we have
\Beq
 \Xi\sol [\Xi\Cau]|_{[-\theta,\theta]}\in
 \P(\cH_k\CauV, \br\|\cdot\|/c';\cH^{l,V}([-\theta,\theta]))
\Eeq
with some $\theta>0$. Now, using \CMPref{Prop. 3.3},
\Beq
 \Xi"(\cdot) := \Xi\sol[\Xi'\mx(b)](\cdot-b)
 \in \P(E,p;\cH^{l,V}([b-\theta,b+\theta]))
\Eeq
is a solution power series, and its Cauchy data at time $b$ agree with that of
$\Xi'\mx$. Hence it can be spliced (cf. Cor. \ref{SpliceLem})
with $\Xi'\mx$ to another solution power series
which shows that $\Xi'\mx$ was not maximal.

Likewise, one shows that $I\mx\not\owns\inf_{t\in I\mx} t$, which yields the
assertion.

Ad (ii): Assume that $b<\infty$, and that there exist $\epsilon>0$, $c'>0$,
and $b'\in I\mx$ such that $\|\Xi'\mx(t)\|_{d/2+\epsilon} <c'$
for all $t\in I\mx\cap [b',\infty)$. Again, we have
\Beq
 \Xi\sol [\Xi\Cau]|_{[-\theta,\theta]}\in
 \P(\cH_{d/2+\epsilon}\CauV, \br\|\cdot\|/c';
 \cH^{l,V}_{d/2+\epsilon}([-\theta,\theta]))
\Eeq
with some $\theta>0$. Now, choosing some $t\in I\mx$ with $t>b'$ and
$t+\theta>b$ and applying the splicing technique once again, we get,
using also the previous Lemma, again a contradiction.
\end{proof}

It is sensible to call $I\mx$ the {\em lifetime interval} of the Cauchy data
$\Xi\CauP$. It follows from assertion (ii) that
the lifetime does not depend on the choice of $k>d/2$ as long as
\eqref{TheLtCauD} holds. However, it depends on $c$; we indicate this
notationally by writing $\Xi'\mxc$, $I\mxc$. What happens if we allow
$c$ to vary?

\begin{prp}\label{Lt4PEH}
Let be given a $\mathbb Z_2$-graded lcs $E$ and an element
\Beqn TheLtCauDMod
 \Xi\CauP\in\P(E;\cH\CauV_k)\sevR.
\Eeq
where $k>d/2$.

(i) There exists a connected open subset $I\mx\seq\mathbb R$ with $I\mx\owns 0$
and a uniquely determined solution power series
$\Xi'\mx\in\P(E;\cH_k^{0,V}(I\mx))\sevR$
which is maximal with the given Cauchy data \eqref{TheLtCauDMod}
(in the analogous sense to that of Prop. \ref{Lt4PEcH}.(i)).

(ii) In case $E$ is a $\mathbb Z_2$-graded Banach space we have
$I\mx= \bigcup_{c>0} I\mxc$ where $I\mxc$ is the lifetime interval
of \eqref{TheLtCauDMod} viewed as element \eqref{TheLtCauD}.

(iii) For arbitrary $E$, $I\mx= \bigcap_p I\mxp$,
where the intersection runs over all those $p\in\CS(E)$ for which
$\Xi\CauP\in\P(\hat E_p;\cH\CauV_k)$, and
$I\mxp$ is the lifetime interval of that element.

(iv) If $a:=\inf_{t\in I\mx} t >-\infty$ then, for any $\epsilon>0$
\Beq
 \operatornamewithlimits{lim\ sup}_{t\to a}
 \|\Xi'\mx[0](t)\|_{\cH\CauV_{d/2+\epsilon}} =\infty
\Eeq
where $\Xi'\mx[0]=(\Xi'\mx)_{(0|0)}$ is the absolute term of $\Xi'\mx$;
likewise for $b:=\sup_{t\in I\mx} t \allowbreak <\infty$.

(v) $I\mx$  is equal to the lifetime interval of
\Beqn OrdCD
 \Xi\CauP[0]\in(\cH\CauV_k)\seven=\P({\mathbf0},0;\cH\CauV_k)\sevR
\Eeq
in the sense of Prop. \ref{Lt4PEcH}.
Here ${\mathbf0}$ is the Banach space consisting of zero alone.
\end{prp}

Again, the lifetime interval does not depend on the choice of $k>d/2$ as
long as \eqref{TheLtCauDMod} holds.

\begin{proof}
The proof is quite analogous to that of the preceding Proposition, using
again the material of \CMPref{3.3}.
\end{proof}

\Brm
(1) \eqref{OrdCD} encodes Cauchy data for the bosonic field equations in the
ordinary, non-super sense. Thus, (v) says roughly that the full field
equations are solvable as long as the underlying bosonic equations are
solvable.

(2) Even for Banach $E$, there is no guaranty that
the power series $\Xi'\mx$ is the Taylor series of a morphism
$\Xi'\mx: U\to\L(\cH_k^{0,V}(I\mx))$ with some neighbourhood
$U\seq\L(E)$ of zero.

(3) We could try to ascend from power series to genuine families of
Cauchy data and solutions. However, since our primary interest is not the
Sobolev quality but the quality $\cEc$, we will do so only in
Thm. \ref{cEcFit}.

(4) For every $k>d/2$, the function
\Beq
 l_+: (\cH_k\CauV)\seven \to \mathbb R_+\cup\infty
\Eeq
which assigns to every $\phi\Cau$ the supremum of its lifetime
interval is lower semicontinous.

Indeed, fix $\phi\Cau$, and let
$s<l_+(\phi\Cau)$. By Prop. \ref{Lt4PEH},
there exists a solution power series
$\Xi'\in\P(\cH\CauV_k,\delta;\cH_k^{0,V}([0,s]))$ with
$\Xi'[0](0)=\phi\Cau$, $\delta>0$.
It follows that $l_+$ is $>s$ in the $\delta$-neighbourhood of
$\phi\Cau$, proving our claim.
\Erm

\subsection{Long-time analyticity}

Avoiding the notion of lifetime interval, the contents of Prop.
\ref{Lt4PEH}.(v)
can be rephrased by saying that a trajectory can serve as a "staircase"
for showing long-time analyticity
in a small neighbourhood of its Cauchy data.

\begin{thm}\label{XiSolXl}
Let be given a trajectory $\phi\in \cH^{0,V}_k(I)\seven$
with $I=[-\theta',\theta]$, $0\le\theta',\theta <\infty$, $0<\theta'+\theta$,
and let $\CData:=\phi(0)$ be its Cauchy data.
Then there exists a closed interval $I'$  with
$I\seq (I')^o$ and a solution power series
\Beq
 \XiCData= \XiCData[\Xi\Cau] \in\P(\cH\CauV_k; \cH^{0,V}_k(I'))\sevR
\Eeq
such that
\Beqn CharXiCData
  \XiCData[\Xi\Cau](0) = \Xi\Cau+\phi(0).
\Eeq
\end{thm}

\Brm
(1) Recalling that $\phi$ is uniquely determined by its Cauchy data,
the notation is sufficiently correct.
It is also introduced to get notational coherence with
\CMPref{3.5}: in case of a complete \sofe., we will construct
in Thm. \ref{MainThm} a superfunctional $\Xi\sol$ the Taylor expansions
of which will be the elements $\XiCData$.

(2) For small times $t$,\ $\XiCData|_{[-t,t]\cap I}$ is simply the translation
(cf. \CMPTransl)
\Beqn TranslXi
 \t\dn{\CData}\br(\Xi\sol|_{[-t,t]\cap I})
\Eeq
of $\Xi\sol$ by the Cauchy data of $\phi$; this is for sufficiently small $t$
well-defined due to Thm. \ref{ShortAnal}.
Thus, $\XiCData$ is a prolongation of \eqref{TranslXi}
to the whole time definition interval of $\phi$ (and, in fact, some $\delta$
beyond). Because of Thm. \ref{UniqCor}.(ii), the
absolute term $\XiCData[0|0] =\phi$ of $\XiCData$ is just the trajectory given.
\Erm

Although this Theorem is a Corollary of Prop. \ref{Lt4PEH}, we give also a
direct proof:

\begin{proof}
We prove this assuming $\theta'=0<\theta$; the case $0=\theta<\theta'$
is handled quite analogously, and the general case follows using
Cor. \ref{SpliceLem}. Let
$c:= 1+ \|\phi\|_{\cH_k^{0,V}(I)}$.
By Thm. \ref{ShortAnal}, we can find some integer $m>0$
such that
\Beq
 \Xi\sol[\Xi\Cau]|_{[-\theta/m,\theta/m]}\in
 \P(\cH_k\CauV,\br\|\cdot\|/c; \cH_k^{0,V}([-\theta/m,\theta/m])).
\Eeq
Define for $i:=0,\dots,m-1$ recursively
\Beq
 \Xi^{(i)} \in\P(\cH_k\CauV; \cH_k^{0,V}([(i-1)\theta/m,(i+1)\theta/m]))
\Eeq
as follows:
\Beq
 \Xi^{(0)}[\Xi\Cau]
 := \t\dn{\CData}\br(\Xi\sol|_{[-\theta/m,\theta/m]})[\Xi\Cau]
  = \Xi\sol[\phi\Cau+\Xi\Cau],
\Eeq
\Beq
 \Xi^{(i)}[\Xi\Cau](t)
 :=\Xi\sol\br[{\Xi^{(i-1)}[\Xi\Cau]((i-1)\theta/m)}](t-(i-1)\theta/m)
  \quad\text{for $i\ge1$.}
\Eeq
However, we have to show that these insertions are legal. The crucial point
is to show by induction, using Thm. \ref{UniqCor}.(ii), that
\Beq
 \Xi^{(i-1)}[0]((i-1)\theta/m)=\phi((i-1)\theta/m)
\Eeq
for $i\ge1$; the legality now follows from \CMPref{3.3}.

Now it follows from Thm. \ref{UniqCor}.(ii) again that the
$\Xi^{(i)}$'s agree on their temporal overlaps,
\Beq
 \Xi^{(i)}|_{[i\theta/m,(i+1)\theta/m]}
 = \Xi^{(i+1)}|_{[i\theta/m,(i+1)\theta/m]}
 \quad
 \text{within\ $\P(\cH_k\CauV; \cH_k^{0,V}([i\theta/m,(i+1)\theta/m]))$.}
\Eeq
By Cor. \ref{SpliceLem}, we can glue these elements together to an
element $\XiCData$ with
\Beq
 \XiCData|_{[(i-1)\theta/m,(i+1)\theta/m]} = \Xi^{(i)}.
\Eeq
Setting $I':= [-\theta/m,\theta(1+1/m)]$, one checks that all requirements
are satisfied.
\end{proof}

In particular, the Theorem always applies to the trivial trajectory $\phi=0$:

\begin{cor}\label{TrivTraj}
For any $\theta>0$ and $k,l$ with $k>\smloss l + d/2$,
\Beq
 \Xi\sol|_{[-\theta,\theta]}\in \P(\cH_k\CauV; \cH_k^{l,V}([-\theta,\theta])).
\Eeq
\qed\end{cor}

\Brm
(1) Thus, for an arbitrary long finite time interval
$[-\theta,\theta]$, \ \ $\Xi\sol|_{[-\theta,\theta]}$ is analytic
for sufficiently small Cauchy data.
Note, however, that there need not be a common domain for the
Cauchy data on which $\Xi\sol(t)$ is analytic for all times.

(2) Thus, even for \sofe.'s which are not complete in the sense of
\ref{Cmplness} below, there is still a one-parameter group of time
evolution which acts on the smf germ $(M\Cau,0)$.

(3) The result could also be proved directly by modifying the proof of
Thm. \ref{ShortAnal}. Also, one needs there only a non-vanishing
convergence radius of $\Delta$, not its entireness.

(4) Using Thm. \ref{UniqCor}.(ii) we get the group property of the
formal solution: For $s,t>0$ we have
\Beq
 \Xi\sol[\Xi\Cau](t) = \Xi\sol[\Xi\sol[\Xi\Cau](s)](t-s).
\Eeq

(5) At this stage, we could already construct a solution smf
$M\sol_{H_k}$ within $\L(C(\mathbb R,\allowbreak\cH_k^{0,V}(\rdm)))$ for each
$k>d/2$.
However, this would not be very useful since this configuration smf is
not Poincar\'e invariant. It cannot be excluded that $M\sol_{H_k}$ is
nevertheless Poincar\'e invariant but we cannot prove this. Therefore we are
using the Sobolev quality only
for intermediate steps, and in the end we are interested in the qualities
$\cE$ and $\cEc$, which are Poincar\'e invariant.
\Erm

\section{Causality and the supermanifold of classical solutions}
\subsection{Spaces of smooth functions}\label{SpaceCic}
In this section, we study the consequences of finite propagation speed, as
it holds in classical field theories.

For the quality of the Cauchy data on the initial hyperplane we choose
the test function space $\cD(\rdm)$; this gives us a maximal reservoir of
superfunctions.
(Of course, with this choice we might miss some interesting classical
solutions; but, at any rate, we come locally arbitrary close to them, and
we avoid quite a lot of technical and rhetorical difficulties.)
We now need a function space on $\rdmm$ such that $\cD(\rdm)$ is the
corresponding space of Cauchy data.

The most naive choice $\cD(\rdmm)$ of compactly supported functions is
not suitable since it contains no nontrivial solutions of the field
equations. On the other hand, if we require from $f\in C^\infty(\rdmm)$
that it have on every time slice compact support then the resulting
space is not Poincar\'e invariant. However, if we additionally require
these supports to grow with time maximally with light velocity
then everything works. (Cf. also \ref{SpaceCit} for a variation of this idea.)

Thus, for $r\ge0$, let
\Beq
 \bV_r:=\{(t,x)\in\rdmm:\quad \br|x|\le r+\br|t|\},
\Eeq
and let temporarily $C^\infty_{\bV_r}(\rdmm)$ be the closed subspace of
$C^\infty(\rdmm)$ which consists of all those elements which have support
in $\bV_r$. Set
\Beq
 \cEc = \bigcup_{r>0} C^\infty_{\bV_r}(\rdmm)
\Eeq
and equip it with the inductive limit topology. This is a strict inductive
limes of Fr\`echet spaces, and hence complete. Also, $\cD(\rdmm)$ is dense
in $\cEc$; hence $\cEc$ is admissible in the sense
of \CMPDefAdm. Moreover, one easily shows that
the subspace $\cEc$ of $C^\infty(\rdmm)$ is invariant under the standard
action of the Poincar\'e group $\Poinc$,
and that the arising action $\Poinc\times\cEc\to\cEc$ is continuous.

For later use, we need a technical notion: Given a seminorm
$p\in\CS(\cD(\rdmm))$, we define the {\em support of $p$}, denoted by $\supp
p$,
as the complement of the set of all $x$ which have a neighbourhood
$U\owns x$ such that $\supp\varphi\seq U$ implies $p(\varphi)=0$.
Obviously, $\supp p$ is closed; using partitions of unity one shows that
$\supp\varphi\seq \rdmm\setminus\supp p$ implies $p(\varphi)=0$.

For every $p\in\CS(C^\infty(\rdmm))$, $\supp p$ is compact (where we
have silently restricted $p$ to $\cD(\rdmm)$). On the other hand:

\begin{lem}\label{CSofCic}
Given $p\in\CS(\cEc)$, \ \ $\supp p \cap \bV_r$ is compact for all $r\ge0$.
\qed\end{lem}

We set
\Beq
 \cEc\Cau := \cD(\rdm),\quad \cEc\CauV := \cD(\rdm,V),\quad
 \cEc\V :=\cEc\otimes V;
\Eeq
thus, the smf's of Cauchy data and of configurations,
\Beq
 M\Cau = \L(\cEc\CauV),\quad M\cfg = \L(\cEc\V)
\Eeq
are now well-defined. Analogously, we set $\cE:=C^\infty(\rdmm)$ and
\Beq
 \cE\Cau :=C^\infty(\rdm),\quad \cE\CauV := C^\infty(\rdm,V),\quad
 \cE\V :=C^\infty(\rdmm,V).
\Eeq

\subsection{Causality}
We call the \sofe. under consideration {\em causal} iff
the function $\hat A$ defined in \eqref{DefHatA} satisfies the following
estimate: for any $\epsilon>0$ there exists $C_\epsilon>0$ such that
\Beqn TheCausEsts
 \br\| \hat A(t,p+\i y) \| \le C_\epsilon \exp((1+\epsilon)\br|yt|),.
\Eeq
for all $t\in\mathbb R$, \ \ $p,y\in\rdm$.

For $(s,x),(t,y)\in\rdmm$ we will write $(s,x)\le(t,y)$
and $(t,y)\ge(s,x)$ iff $(t,y)$ lies in the forward light cone of
$(s,x)$, i.e. $(t-s)^2\ge(y-x)^2$.

\begin{lem}\label{CauUniq}
Suppose that the \sofe. is causal.

(i) We have
\Beq
 \supp A \seq \{(t,x)\in\mathbb R\times\rdm:\ \  \br|x|\le |t|\},
\Eeq
\Bmln GCausal
 \supp G(t,s,x) = \{ (t,s,x) \in\mathbb R\times\mathbb R\times\rdm: \\
    \text{($s\ge0$ and $0\le(t-s,y-x)$) or ($s\le0$ and $0\ge(t-s,y-x)$)}\}.
\Eml

(ii) The assignment $(\xi\Cau,g)\mapsto\A\xi\Cau + \G g$
of Lemma \ref{GreensAndProps} restricts to a continuous linear map
\Beq
 \cEc\CauV\oplus\cEc\V\to\cEc\V.
\Eeq

(iii) Let $p=(s',x')\in\rdmm$ be a point with $s'\not=0$, and set
\Baln Omega
 &\Omega(p) =
 \begin{cases}
  \{(s,x)\in\rdmm:\ (s,x) < (s',x'),\ 0 < s\}  & \text{if $s'>0$,}\\
  \{(s,x)\in\rdmm:\ (s,x) > (s',x'),\ 0 > s\}  & \text{if $s'<0$,}
 \end{cases}
 \\
 \label{AndJ}
 &\J(p) = \{x\in\rdm:\ \br|x-x'|<\br|s'|\}.
\Eal
Also, let $I=[0,s']$, and let be given a formal power series
$\Xi'\in\P_f(E;\cH_k^{0,V}(I))$ with $k>d/2$ and some $\mathbb Z_2$-lcs $E$
which satisfies
\Beq
 (\partial_t+K(\partial_x))\Xi'|_{\Omega(p)} = 0,\quad
 \Xi'(0)|_{\J(p)}=0.
\Eeq
Then $\Xi'(p)=0$.
\end{lem}

\begin{proof}
(i) is an immediate consequence of the Paley-Wiener theorem while (ii)
follows by standard techniques.

Ad (iii). By standard techniques (cf. e.~g. the proof of
Thm. \ref{CausUniqThm} for one possibility), one proves this for ordinary
functions; then one looks at the coefficient functions of $\Xi'$.
\end{proof}

\subsection{Solution families in the causal case}\label{CausFam}
We will call any element $\Xi'\in\M^{\cEc\V}(Z)$
a {\em $Z$-family of quality $\cEc$}.
Because of the inclusion $\cEc\V\seq\cH^{l,V}_k(\mathbb R)$,
such an element can be viewed as family of quality $\cH^l_k$
with time definition domain $\mathbb R$ for all $k,l$ with $k>d/2+\smloss l$.

On the other hand, we will need also {\em families of quality $\cE$}, i.~e.
elements $\Xi'\in\M^{\cE\V}(Z)$; the notions {\em family of solutions}
and {\em pullback} make still obvious sense for them.

One family of quality $\cEc$ is given a priori, namely the $M\cfg$-family
\Beq
 \Xi=(\Phi|\Psi)\in\M^{\cEc\V}(M\cfg)
\Eeq
where, we recall, $M\cfg=\L(\cEc\V)$ is the smf of
configurations of quality $\cEc$, and $\Xi$ is the standard coordinate
(cf. \whatref{\SFctAnFu}).

$\Xi$ is in fact the {\em universal family of quality $\cEc$}:
Given an arbitrary $Z$-family $\Xi'$ of quality $\cEc$, it defines by
Lemma \ref{CoordLem}
a {\em classifying morphism}
\Beq
 \check{\Xi'}:Z\to M\cfg, \quad \widehat{\check{\Xi'}} = \Xi'
\Eeq
and $\Xi'$ arises from $\Xi$ just by pullback: $\Xi'=\check{\Xi'}^*(\Xi)$.

\Brm (1) In the language of category theory, this means that the
cofunctor
\Beq
 \{\text{supermanifolds}\}\to\{\text{sets}\},\qquad Z\mapsto\M^{\cEc}(Z),
\Eeq
is represented by the object $M\cfg$ with the universal element $\Xi$.

(2) Of course, universal families exist also for other time definition
domains and qualities: One simply takes functional coordinates on
the linear supermanifolds over the corresponding locally convex
function spaces. However, we have no use for them.
\Erm

For a superfunction with values in continuous functions on $\rdmm$,
i.~e. $K\in\O^{C(\rdmm)}(Z)$, let the {\em target support of $K$} be
defined as
\Beq
 \opn t-supp K := \opn Closure \Bigl(\{x\in\rdmm:\ \ K(x) \not= 0\}\Bigr),
\Eeq
where, of course, $K(x) = \delta_x\circ K$. This should not be confused
with the support of a power series as defined in \CMPref{3.11}.

For a causal \sofe., we have the following strengthening of
Thm. \ref{UniqCor}:

\begin{thm}\label{CausUniqThm}
Suppose that the \sofe. is causal.
Let be given a point $p=(s',x')\in\rdmm$ with $s'\not=0$,  and let
$\Omega(p), \J(p)$ be as in \eqref{Omega}, \eqref{AndJ}. Also, let
$I=[0,s']$ if $s'>0$ and $I=[s',0]$ if $s'<0$, respectively.

(i) Let be given a $Z$-family $\Xi'\in\M^{\cH_k^{l,V}(I)}(Z)$
of configurations with $k>d/2$, and suppose that
\Beqn SolOnOmega
 L[\Xi']|_{\Omega(p)} =0
\Eeq
within $\M^{\cD'(\Omega(p))\otimes V}(Z)$ (i. e., loosely said, $\Xi'$
"is a solution on the open space-time domain $\Omega(p)$").
Then $\Xi'$ satisfies the integral equation
\Beqn TYFamIntEq
 \Xi'(t,y) = \Xi\free[\Xi'(0)](t,y) + \G\Delta[\Xi'](t,y)
\Eeq
within $\O\V(Z)$ for all $(t,y)\in\Omega(p)$.

(ii) Let be given two $Z$-families $\Xi',\Xi"\in\M^{\cH_k^{0,V}(I)}(Z)$
with $k>d/2$, and suppose that
\Beq
 L[\Xi']|_{\Omega(p)} = L[\Xi"]|_{\Omega(p)} =0,\quad
 \left(\Xi'(0)-\Xi"(0)\right)|_{\J(p)}=0.
\Eeq
Then $\br(\Xi'-\Xi")|_{\Omega(p)}=0$.

(iii) Suppose that for some $r>0$
\Beq
 \opn t-supp \Xi'(0) \seq \ball r^d,
\Eeq
i. e. $\Xi'(0)|_{\rdm\setminus \ball r^d} = 0$.
Then
\Beq
 \opn t-supp \Xi'(t) \seq \ball{(r+\br|t|)}^d
\Eeq
for all $t\in I$.
\end{thm}

\begin{proof} Ad (i). Set temporarily
$F:=\Xi' - \Xi\free[\Xi'(0)] - \G\Delta[\Xi']\in\O^{\cH_k^{l,V}(I)}(Z)$.
We find for $(t,y)\in\Omega(p)$, using \eqref{SolOnOmega},
\Beq
 (\partial_t + K(\partial_x))F(t,y) =
 - \Delta[\Xi'](t,y) + (\partial_t + K(\partial_x))\G\Delta[\Xi'](t,y).
\Eeq
Using \eqref{KOnG}, this vanishes. By Lemma \ref{CauUniq}.(iii), $F(t,y)=0$.

Ad (ii). We may suppose $Z$ to be a superdomain
$Z\seq\L(E)$; pick one $z\in Z$. Now we may choose some $r\in\CS(E)$ such that
the relevant Taylor expansions $\Xi'_z,\Xi"_z$ lie in the Banach space
$\P(E,r;\cH_k^{0,V}(I))$.

After a spatial translation, we may assume $x'=0$, so that
the closure of $\J(p)$ is the closed ball $\ball{s'}^d$.
Also, we may assume $s'>0$; otherwise, the following arguments have
to be "mirrored".

Suppose there exists some $t_1\in I$
with $(\Xi'_z-\Xi")_z(t_1)|_{\ball{(s'-t_1)}^d}\not=0$. Now the set
$\{t\in[0,t_1]:\ (\Xi'-\Xi")_z(t)|_{\ball{(s'-t)}^d} =0\}$ is easily seen
to be closed; let $t_2$ be its maximum.
By passing to the shifted families $\Xi'_z(\cdot-t_2)$,
$\Xi"_z(\cdot-t_2)$, we may assume $t_2=0$.

{}From \eqref{TYFamIntEq} and the hypotheses we get with $\Theta:=\Xi"-\Xi'$
that
\Beqn IntEqTheta
 \Theta_z(t,y) = \G(\Delta[\Xi'_z+ \Theta_z]- \Delta[\Xi'_z])(t,y)
\Eeq
for $(t,y)\in\Omega(p)$ where the integral over $s$ runs effectively
over $[0,t]$.

Our problem is that \eqref{IntEqTheta} does not hold for all $(t,y)$.
Therefore, we have to use temporarily the standard Sobolev space on
the closed ball $\ball c^d$: For integer $k>d/2$, let
\Beq
 H_k(\ball c^d) := \Bigl\{ f\in L_2(\rdm):\quad \supp f \seq \ball c^d,\
 \br\|f\| := \sum\nolimits_{\nu\in\mathbb Z_+^d,\ \br|\nu|\le k}
  \br\|\partial^\nu f\|_{L_2} < \infty \Bigr\};
\Eeq
for non-integer $k>d/2$, define $H_k(\ball c^d)$ by interpolation.
It is well-known (cf. e.~g. \cite{[Taylor]}) that there exists
bounded linear operators $\E_c: H_k(\ball c^d) \to H_k(\rdm)$
which are right inverses to restriction, i.~e. $\E_c(f)|_{\ball c^d}=f$.
In fact, having choosen $\E_1$, we may and will set
$\E_c(f)(x):=  \E_1(f(c\cdot))(x/c)$.
Now these operators yield a bounded linear operator
\Beq
 \E: \cH^{0,V}_k([0,\frac {s'}2]) \to \cH^{0,V}_k([0,\frac {s'2}2]),\quad
 \E(f)(t,x) := \E_{s'-t}(f(t,\cdot))(x);
\Eeq
thus, $\E(f)$ depends only on the restriction $f|_{\Omega(p)}$ (of
course, the role of $s'/2$ could be played by any number in $(0,s')$).

Now, using the support property \eqref{GCausal} of the Green functions
we get from \eqref{IntEqTheta}
\Beqn Theta=F
 \Theta_z|_{\Omega(p)} =
 \G(\Delta[\Xi'_z+ \E(\Theta_z)] - \Delta[\Xi'_z])|_{\Omega(p)}.
\Eeq
For shortness, we will write again $\br\|\cdot\|_F$ for the norms in
$\P(E,r;F)$ where $F$ is one of the Banach spaces $\cH^l_k(I)$,
$\cH_k\CauV$.

Using Lemma \ref{GreensAndProps}.(iii) we get that
there exists a constant $C_1>0$ such that
\Beq
 \br\|\G(\Delta[\Xi'_z+ \E(\Theta_z)] - \Delta[\Xi'_z])(t)\|_{\cH_k\CauV}
 \le C_1 \br|t|
 \br\| \Delta[\Xi'_z+ \E(\Theta_z)] - \Delta[\Xi'_z] \|_{\cH_k^{0,V}([0,t])}
\Eeq
for $t\in [0,s'/2]$. Because of \eqref{Theta=F}, we have a fortiori,
\Beq
  \br\|\Theta_z(t)\|_{H_{k+1}(\ball{(s'-t)}^d)} \le C_1 \br|t|
 \br\| \Delta[\Xi'_z+ \E(\Theta_z)] - \Delta[\Xi'_z] \|_{\cH_k^{0,V}([0,t])};
\Eeq
because of the continuity of $\E$ this implies
\Beqn ContOfE
 \br\|\E(\Theta_z(t))\|_{\cH_k\CauV}
 \le C_2 \br|t|
 \br\| \Delta[\Xi'_z+ \E(\Theta_z)] - \Delta[\Xi'_z] \|
\Eeq
with some $C_2>0$. To estimate the \rhs\ of this, we introduce
temporarily new indeterminates $\ul\Theta_i$, with $i=1,\dots,N$, and
$|\ul\Theta_i|=|\ul\Xi_i|$. Working in the power series algebra
$\mathbb C[[\partial^*\ul\Xi,\partial^*\ul\Theta]]$, we can expand
\Beq
 \Delta_j[\ul\Xi+\ul\Theta] - \Delta_j[\ul\Xi] =
 \sum_{k=1}^N \sum_{\br|\nu|\le \sd_k-\sd(\Delta_j)}
  B_{\nu,jk}[\ul\Xi]\partial^\nu \ul\Theta_k
 +  R_j[\ul\Xi,\ul\Theta]
\Eeq
where both $B_{\nu,jk}[\ul\Xi],\ R_j[\ul\Xi,\ul\Theta]$ are entire, and
$R_j[\ul\Xi,\ul\Theta]$ is in $\ul\Theta$ of lower degree $\ge2$.
Using also Lemma \ref{PSEst} we get that there exists $C_3>0$ with
\Beqn EstDiffDelta
 \br\| \Delta[\Xi'_z+ \E(\Theta_z)] - \Delta[\Xi'_z]\| \le
 C_3 \br\|\E(\Theta_z)\|
\Eeq
(both norms in $\P(E,r;\cH_k^{0,V}([0,t]))$) for $t\in [0,s'/2]$.
Putting \eqref{ContOfE}, \eqref{EstDiffDelta} together we get
\Beq
 \br\|\E(\Theta_z(t))\|_{\cH_k\CauV} \le C_4 \br|t|
 \br\|\E(\Theta_z|_{[0,t]})\|_{\cH_k^{0,V}([0,t])}
\Eeq
for $t\in [0,s'/2]$ with $C_4:=C_2C_3$.
Now, for (say) $t < 1/(2C_4)$, this estimate implies
$\br\|\E(\Theta_z)|_{[0,t]}\|=0$, which yields a contradiction
to our assumptions.

Ad (iii). This follows because $\Xi":=0$ is a solution family.
\end{proof}

\begin{lem}\label{SmCSols}
Suppose that the \sofe. is causal.
Let be given a trajectory $\phi\in\cH_k^{0,V}(I)\seven$ such that
$\phi(0)\in(\cEc\CauV)\seven$.

(i) We have
\Beqn AutSm
 \phi\in C^\infty(I\times\rdm)\otimes V\seven.
\Eeq
Moreover, if $\supp \phi(0) \seq \ball r^d$ for some $r>0$
then
\Beqn SuppSpr
 \supp\phi(t) \seq \ball{(r+\br|t|)}^d
\Eeq
for all $t\in I$.

(ii) In particular, if $I=\mathbb R$ then $\phi\in(\cEc\V)\seven$.
\end{lem}

\begin{proof}
Ad (i).
By Prop. \ref{Lt4PEH} and Prop. \ref{TempSmooth}, we have
$\phi\in\cH_{k'}^{l,V}(I)\seven$ for all $k',l$ with $k'>\smloss l+d/2$.
On the other hand, by Lemma \ref{HklLemma}.(ii), we have a continuous embedding
$\bigcap_{k,l:\ k>\smloss l+d/2}\cH^{l,V}_k(I)\seq C^\infty(I\times\rdm)\otimes
V$,
which proves \eqref{AutSm}.
\eqref{SuppSpr} is a special case of Thm. \ref{UniqCor}.(iii).

Ad (ii). Obvious.
\end{proof}

\subsection{Analyticity with targets $\protect{\cE}$ and $\protect{\cEc}$}
\label{BigTrick}

Causality will provide the deux ex machina, which allows to conclude
from Sobolev continuity to continuity in the quite different topologies
of $\cE$ and $\cEc$.

We begin with some technical preparations.
Given a bounded open set $\Omega\Subset\rdmm$, we denote by
$\J(\Omega)\Subset\rdm$ the {\em causal influence domain of $\Omega$}
on the Cauchy hyperplane, i.~e. the set of all $x\in\rdm$ such that
$(0,x)$ lies in the twosided light cone of a point in $\Omega$.

For $\Omega\Subset\rdmm$, $l\ge0$, define the seminorm
$q_{l,\Omega}\in\CS(\cE\V)$ by
\Beq
 q_{l,\Omega}(\xi) = \sum_{i=1}^N
 \sup_{(t,x)\in\Omega} \sum_{\nu\in\mathbb Z_+^{d+1},\ |\nu|\le l}
 \br|\partial^\nu \xi_i(t,x)|;
\Eeq
thus $\supp q = \Omega$ (cf. \ref{SpaceCic}).

For $J\Subset\rdm$, $k\ge0$, define the seminorm $p_{k,J}\in\CS(\cE\CauV)$ by
\Beq
 p_{k,J}(\xi\Cau) := \sum_{i=1}^N \sup_{x\in J}
 \sum_{\nu\in\mathbb Z_+^d,\ |\nu|\le k} \br|\partial^\nu \xi\Cau_i(x)|;
\Eeq
thus, $\supp p_{k,J}=J$.

\begin{lem}\label{CausInfEst}
Suppose that the \sofe. is causal. Fix Cauchy data
$\CData\in(\cEc\CauV)\seven$
the lifetime interval of which is the whole time axis $\mathbb R$, so that by
Lemma \ref{SmCSols}, there exists an all-time trajectory
$\phi\in(\cEc\V)\seven$ with these Cauchy data.

(i) For $\Omega\Subset\rdmm$, $l\ge0$, let
$k>\smloss l+d/2 + \max\{\sd_1,\dots,\sd_N\}$.
Then, for all $\epsilon>0$,
the power series $\XiCData[\Xi\Cau]$
given by Thm. \ref{XiSolXl} satisfies a
$(q_{l,\Omega}, C_\epsilon p_{k,J_\epsilon})$-estimate
(cf. \CMPref{3.1}) with some $C_\epsilon>0$, where
$J_\epsilon=U_\epsilon(\J(\Omega))$ is the
$\epsilon$-neighbourhood of $\J(\Omega)$.

(ii) Let $q\in\CS(\cE\V)$ be arbitrary. Then there exists $k>0$ such that
for all $\epsilon>0$, $\XiCData[\Xi\Cau]$ satisfies the
$(q,C_\epsilon p_{k,J_\epsilon})$-estimate with some $C_\epsilon>0$,
where $J_\epsilon=U_\epsilon(\J(\supp q))$.
\end{lem}

\begin{proof}
Ad (i). Let $I\seq\mathbb R$ be the projection of $\Omega$ onto the time axis.
By Lemma \ref{HklLemma}.(ii), there exists a constant $C_1$ such that
\Beq
 q_{l,\Omega}(\varphi) \le  C_1 \cdot\br\|\varphi|_I\|_{\cH_k^l(I)}
\Eeq
for $\phi\in\cH_k^l(I)$.
Combining this with the Sobolev analyticity of
$\XiCData[\Xi\Cau]$ given by Thm. \ref{XiSolXl}, there exists a
constant $C_2$ such that we have for $r,s\ge0$,
$\varphi^1,\dots,\varphi^r\in(\cEc\CauV)\seven$,
$\psi^1,\dots,\psi^s\in(\cEc\CauV)\sodd$
\Beq
 q_{l,\Omega}\Bigl(\Bigl<\br(\XiCData)_{r|s},
  \bigotimes_{m=1}^r \varphi^m\otimes \bigotimes_{n=1}^s \Pi \psi^n\Bigl>\Bigr)
 \le
 C_2\cdot\prod_{m=1}^r \br\|\varphi^m\|_{\cH\CauV_k}\cdot
  \prod_{n=1}^s\br\|\psi^n\|_{\cH\CauV_k}
\Eeq
(cf. \CMPref{3.1} for the notation on the \lhs).
Now choose some buffer function $h\in\cD'(\rdm)$ with $\supp h\seq J_\epsilon$
and $h|_{\J(\Omega)}=1$. By causality (cf. Thm. \ref{CausUniqThm}.(ii)),
we have $\XiCData[\Xi\Cau]|_\Omega = \XiCData[h\Xi\Cau]|_\Omega$, and hence
\Bal
 q_{l,\Omega}\Bigl(\Bigl<\br(\XiCData)_{r|s},
 \bigotimes_{m=1}^r \varphi^m\otimes
 \bigotimes_{n=1}^s \Pi \psi^n\Bigr>\Bigr)
 &=
 q_{l,\Omega}\Bigl(\Bigl<\br(\XiCData)_{r|s},
 \bigotimes_{m=1}^r (h\varphi^m)\otimes
 \bigotimes_{n=1}^s \Pi(h\psi^n)\Bigr>\Bigr)\\
 &\le
 C_2\cdot\prod_{m=1}^r
 \br\|h\varphi^m\|_{\cH\CauV_k}\cdot \prod_{n=1}^s\br\|h\psi^n\|_{\cH\CauV_k}
\Eal
But obviously $\br\|h\cdot\|_{\cH\CauV_k}$ is estimated from above by
$C_\epsilon p_{k,J_\epsilon}(\cdot)$ with some $C_\epsilon>0$,
and the assertion follows.

Ad (ii). Since the collection of all $q_{l,\Omega}$ defines the topology
of $\cE\V$, there exist $l,C'$,and $\Omega'\Subset\rdmm$ such that
$q\le C'q_{l,\Omega'}$.
However, $\Omega'$ may be larger than $\supp q$. Choose a buffer
function $g\in\cD(\rdmm)$,\ \  $g\ge0$, with $g|_{\supp q}=1$,\ \
$\supp g\seq J_{\epsilon/2}$. Then
\Beq
 q(\cdot) = q(g\cdot) \le C'q_{l,\Omega'}(g\cdot)
 \le C'_\epsilon q_{l,J_{\epsilon/2}}(\cdot)
\Eeq
with some $C'_\epsilon>0$. The assertion now follows from (i).
\end{proof}

\begin{prp}\label{TargetCE}
Suppose that the \sofe. is causal.
Let be given bosonic Cauchy data $\CData\in(\cEc\CauV)\seven$
the lifetime interval of which is the whole time axis $\mathbb R$.
Then $\XiCData[\Xi\Cau]$ is an analytic power series from
$\cEc\CauV$ to $\cEc\V$:
\Beq
 \XiCData[\Xi\Cau]\in\P(\cEc\CauV; \cEc\V)\sevR.
\Eeq
\end{prp}

\begin{proof}
Let be given a seminorm $q\in\CS(\cEc\V)$. With standard methods one
constructs for $i>0$ buffer functions $f_i\in C^\infty(\rdmm)$
with $f_i|_{\bV_{i-1}}=0$, $f_i|_{\rdmm\setminus\bV_i}=1$ where the
$\bV_i$  are as in \ref{SpaceCic}. Set for convenience $f_0:=1$.
For the seminorms $q_i:=q((f_i - f_{i+1})\cdot)\in\CS(\cEc\V)$ we get
\Beqn pIsSumQ
 q(\varphi) \le \sum_{i\ge0} q_i(\varphi)
\Eeq
for all $\phi\in\cEc\V$, where in fact only finitely many terms on the \rhs\
are non-zero. Now
\Beq
 \supp q_i\seq \bV_{i+1} \cap \supp q
\Eeq
which is by Lemma \ref{CSofCic} compact. Also, for $i\ge1$, we have
$(f_i - f_{i+1})|_{\bV_{i-1}}=0$ and hence
\Beqn OutOfVi-1
 \supp q_i \cap \bV_{i-1}=\emptyset.
\Eeq
Because of \eqref{OutOfVi-1}, we have
$\J(\supp q_i)\seq \{x\in\rdm: \ \ \br\|x\|\ge i-1\}$ for $i\ge1$; hence,
setting $J_i:= \{x\in\rdm: \ \ \br\|x\|\ge i-2\}$,
Lemma \ref{CausInfEst}.(ii) yields for each $i$ numbers
$C_i>0$,\ \ $k_i\ge0$ such that $\XiCData[\Xi\Cau]$ satisfies a
$\br( q_i, C_i p_{k_i,J_i})$-estimate.

It follows that for each $\varphi\in\cEc\CauV$, the sum
\Beq
   p(\varphi) := \sum_i C_i p_{k_i,J_i}(\varphi)
\Eeq
has only finitely many nonvanishing terms; using
\cite[Thm. 15.4.1]{[Hormander]}, we have $p:=p(\cdot)\in\CS(\cEc\CauV)$.
It follows directly from the definition of the $(q,p)$-estimates
(cf. \CMPref{3.1}) and \eqref{pIsSumQ} that the
$\br( q_i, C_i p_{k_i,J_i})$-estimates
for $\XiCData[\Xi\Cau]$ imply the $(q,p)$-estimate wanted.
\end{proof}

\begin{thm}\label{cEcFit}
Suppose that the \sofe. is causal.
Let be given an smf $Z$ and a superfunction $\Xi\CauP\in\M^{\cEc\CauV}(Z)$
(this encodes an smf morphism $(\Xi\CauP)\spcheck: Z\to M\Cau$, i.~e. a
family of Cauchy data).

Suppose that for each $z\in Z$, there exists
a smooth all-time trajectory $\phi_z\in(\cEc\V)\seven$ with
$\phi_z(0)=\Xi\CauP(z)$.
Then there exists a unique $Z$-family of solutions $\Xi'\in\M^{\cEc\V}(Z)$
which has $\Xi\CauP$ as its Cauchy data, i. e. $\Xi'(0) = \Xi\CauP$.
The Taylor expansion of $\Xi'$ at $z$ is given by
\Beqn DefXiZ
 \Xi'_z=\Xi\sol_{\phi_z(0)}[\Xi\CauP_z-\phi_z(0)]
\Eeq
where $\Xi\sol_{\phi_z(0)}$ is given by Thm. \ref{XiSolXl}. Note that the
insertion is defined since the power series inserted has no absolute term.

Also, the underlying map of the arising smf morphism
$\check{\Xi'}:Z\to M\cfg$ is $z\mapsto\phi_z$.
\end{thm}

\begin{proof}
We may assume that $Z\seq\L(E)$ is a superdomain. Using Prop.
\ref{TargetCE}, we get a map
\Beq
 \Space Z\owns z\mapsto\Xi'_z\in\P(E;\cEc\V)
\Eeq
where $\Xi'_z$ is defined by \eqref{DefXiZ}.
We have to show that this is an element of $\M^{\cEc\V}(Z)$.
This task is simplified by remarking that the set of all functionals
\Beqn StrSepSet
 \delta^i_{(t,x)}: \cEc\V\to\mathbb R, \quad \xi=(\xi_i)\mapsto \xi_i(t,x),
\Eeq
with $i=1,\dots,N$ and $(t,x)\in\rdmm$ is strictly separating in the sense
of \whatref{\StrictSep}; it follows that it is sufficient to
check that for each $\delta^i_{(t,x)}$, the assignment
\Beq
 \Space Z \owns z\mapsto
  \delta^i_{(t,x)}\circ\Xi'_z\in\P(E;\mathbb R)
\Eeq
is an element of $\O(Z)$ (note that $\delta^i_{(t,x)}$ is even for
$i\le N\seven$ and odd otherwise). Thus it is sufficient to prove:
Fix $i,(t,x)$ and $z\in Z$. There exists $p\in\CS(E)$ such that
$\delta^i_{(t,x)}\Xi'_z\in\P(E,p;\mathbb R)$, and for
$z'\in Z$,\ \ $p(z'-z)<1$, we have
\Beqn 2Prv
 \t_{z'-z}\delta^i_{(t,x)}\Xi'_z = \delta^i_{(t,x)}\Xi'_{z'}.
\Eeq
Indeed, set (say) $k:=d/2+1$, and $H:=\cH^{0,V}_k([-\br|t|-1,\br|t|+1])$.
Choose $p\in\CS(E)$ such that
$\Xi\CauP_z-\phi_z(0)\in\P(E,p;\cH\CauV_k)$; since there is no absolute
term we may assume by dilating $p$ that $C\br\|\Xi\CauP_z-\phi_z(0)\|<1$.
The composite \eqref{DefXiZ} is now defined in the sense of
\CMPInsMechPrp, and we get $\Xi'_z\in\P(E,p;H)$.

Choose by Thm. \ref{XiSolXl} some $C>0$ such that
\Beq
 \Xi\sol_{\phi_z(0)}[\Xi\Cau]
 \in\P(\cH\CauV_k,C\br\|\cdot\|;H).
\Eeq
For $z'\in Z$,\ \ $p(z'-z)<1$, we have
\Beq
 \Xi\CauP_{z'}=\t_{z'-z}\Xi\CauP_z\in\P(E,cp;H),\quad c:=1-p(z'-z),
\Eeq
and hence
\Beq
 \t_{z'-z}\Xi'_z =\Xi\sol_{\phi_z(0)}[\t_{z'-z}(\Xi\CauP_z-\phi_z(0))]
 = \Xi\sol_{\phi_z(0)} [\Xi\CauP_{z'}-\phi_z(0)] \in\P(E,cp;H);
\Eeq
since translation is an algebra homomorphism,
this element is a solution power series.
Using \eqref{CharXiCData}, we find its Cauchy data as
\Beq
 \t_{z'-z}\Xi'_z(0) = \Xi\CauP_{z'}.
\Eeq
On the other hand, we have also the element $\Xi'_{z'}\in\P(E,c'p;H)$
with some $c'>0$ which yields a solution family $\Xi'_{z'}\in\M^H(V')$ where
$V'$ is the $c'$-fold multiple of the unit ball of $p$ in $\L(E)$. Since
it has the same Cauchy data as $\t_{z'-z}\Xi'_z$, we get from
Thm. \ref{UniqCor}.(ii) that $\t_{z'-z}\Xi'_z = \Xi'_{z'}$ within
$\P(E;H)$. A fortiori, we have \eqref{2Prv}.
\end{proof}

We need the power series $\XiCData[\Xi\Cau]$ also for the case that
$\CData\in\cE\CauV$ has no longer compact support, so that
the proofs of Thm. \ref{XiSolXl} and Prop. \ref{Lt4PEH} break down.

We call a bosonic Cauchy datum $\CData\in(\cE\CauV)\seven$
{\em approximable} if there exists a sequence of compactly
supported bosonic Cauchy data
$\phi_{(n)}\Cau\in(\cEc\CauV)\seven$,\quad $n\in\mathbb Z_+$,
such that $\phi_{(n)}\Cau|_{\ball n^d} = \CData|_{\ball n^d}$
for all $i$, and each $\phi_{(n)}\Cau$ has the whole time axis
as lifetime interval, i. e., there exists an all-time solution
$\phi_{(n)}\in(\cEc\V)\seven$ of the bosonic field equations with these
Cauchy data.

\begin{lem}\label{XiCDWR}
Suppose that the \sofe. is causal.
Given an approximable bosonic Cauchy datum $\CData\in(\cE\CauV)\seven$,
there exists a solution power series
$\XiCData[\Xi\Cau]\in\P(\cE\CauV;\cE\V)\sevR$ such that
\Beqn CDtaWR
  \XiCData[\Xi\Cau](0) = \Xi\Cau+\phi(0).
\Eeq
\end{lem}

\begin{proof}
Composing the power series
$\Xi\sol_{\phi_{(n)}\Cau}\in\P(\cEc\CauV;\cEc\V)$
given by (ii) and Prop. \ref{TargetCE} with the projection
$\cEc\V\to C^\infty(\ball n^{d+1})\otimes V$ we get a sequence of power series
\Beq
 \Xi_{(n)} := \Xi\sol_{\phi_{(n)}\Cau}|_{\ball n^{d+1}}
 \in\P(\cEc\CauV; C^\infty(\ball n^{d+1})\otimes V)\sevR.
\Eeq
Because of Thm. \ref{CausUniqThm}.(ii) and (i), the restrictions of
$\Xi_{(n+1)}$
and $\Xi_{(n)}$ onto $\ball n^{d+1}$ coincide. Hence there exists a power
series $\XiCData[\Xi\Cau]\in\P(\cEc\CauV;\cE\V)$ whose restriction
onto $\ball n^{d+1}$ is $\Xi_{(n)}$.
It is clear that this is a solution power
series which satisfies \eqref{CDtaWR}; the fact that it is actually
analytic with respect to the source space $\cE\CauV$ follows from
Lemma \ref{CausInfEst}.(i) and the construction.
\end{proof}

We also have the analogon of Thm. \ref{cEcFit}:

\begin{thm}\label{cEcFitWR}
Suppose that the \sofe. is causal.
Let be given an smf $Z$ and a superfunction $\Xi\CauP\in\M^{\cE\CauV}(Z)$.
Suppose that for each $z\in Z$, the Cauchy datum
$\phi\CauP(z)\in(\cE\CauV)\seven$ is approximable.

Then there exists a unique $Z$-family of solutions $\Xi'\in\M^{\cE\V}(Z)$
which has $\Xi\CauP$ as its Cauchy datum.
Again, the Taylor expansion of $\Xi'$ at $z$ is given by \eqref{DefXiZ},
where this time, $\Xi\sol_{\phi_z(0)}$ is given by Lemma \ref{XiCDWR}.
\end{thm}

\begin{proof}
Quite analogous to that of Thm. \ref{cEcFit}; instead of the functionals
$\delta^i_{(t,x)}$, one could use also the seminorms
$p_{l,\Omega}$.
\end{proof}

Finally, we will need a variant of Prop. \ref{TargetCE} which describes
compactly supported local excitations around a classical solution:

\begin{prp}\label{ExcAnal}
Suppose that the \sofe. is causal.
Let be given an approximable bosonic Cauchy datum $\CData\in(\cE\CauV)\seven$,
and let $\phi:=\XiCData[(0,0),0]\in (\cE\V)\seven$ be the corresponding
solution. Then the power series
\Beq
 \Xi\exc_\phi[\Xi\Cau]:= \XiCData[\Xi\Cau]-\phi
\Eeq
satisfies $\Xi\exc_\phi\in\P(\cEc\CauV; \cEc\V)\sevR$.
\end{prp}

\begin{proof}
First one shows the analogon of Lemma \ref{CausInfEst} for $\Xi\exc_\phi$;
in the proof, one replaces the Sobolev analyticity of $\XiCData$ by the
fact that $\Xi\exc_\phi\in\P(\cE\CauV; \cE\V)$, and the necessary causality
property is again provided by Thm. \ref{CausUniqThm}.(ii). Having this, the
proof of Prop. \ref{TargetCE} carries over.
\end{proof}


\subsection{Completeness}\label{Cmplness}
Loosely said, we call a \sofe. complete iff the underlying bosonic
\sofe. is globally solvable:

\begin{thm}
For a causal \sofe., the following conditions are equivalent:

(i) For every smooth solution
$\phi\in C^\infty((a,b)\times \mathbb R^d) \otimes V\seven$ of
the underlying bosonic field equations on a bounded open time
interval $(a,b)$ such that $\supp\phi(t)$ is compact for all $t\in(a,b)$
there exists a Sobolev index $k>d/2$ such that
\Beqn APEst
 \sup_{t\in(a,b)} \br\|\phi(t)\|_{\cH\CauV_k} < \infty.
\Eeq

(ii) The underlying bosonic equations are all-time solvable with
quality $\cEc$:

Given bosonic Cauchy data $\phi\Cau\allowbreak \in(\cEc\CauV)\seven$
there exists an element $\phi\in(\cEc\V)\seven$
with these Cauchy data which solves the field equations.

(iii) The underlying bosonic equations are all-time solvable with
quality $\cE$:

Given bosonic Cauchy data $\phi\Cau\allowbreak \in(\cE\CauV)\seven$
there exists an element $\phi\in (\cE\V)\seven$
with these Cauchy data which solves the field equations.

\noindent If these conditions are satisfied we call the \sofe. {\em complete}.
\end{thm}

\Brm
(1) Of course, in case of completeness, \eqref{APEst}
holds for all $k>d/2$ for every $\phi$ as in (i).

(2)
The solutions provided in (iii), (iv) are necessarily uniquely determined.
We need no information about their continuous dependence on the initial
data since our theory yields automatically real-analytic dependence.

(3) It would be nice to add the following conditions to the list:

{\em
(iv) The underlying bosonic equations are all-time solvable with
some Sobolev quality $k>d/2$:

Given bosonic Cauchy data $\phi\Cau \in(\cH\CauV_k)\seven$
there exists an element $\phi\in \cH_k^{0,V}(\mathbb R)\seven$
with these Cauchy data which solves the field equations.

(v) The solvability assertion of (iv) holds for all Sobolev orders
$k>d/2$.
}

However, at the time being, we cannot exclude the possibility that
even for a complete \sofe., there exist bosonic Cauchy
data $\phi\Cau\in(\cH\CauV_k)\seven\setminus(\cE\CauV)\seven$
with finite lifetime $t_0$; for any sequence
$\phi_{(i)}\Cau\in(\cEc\CauV)\seven$ converging to
$\phi\Cau$ within $\cH\CauV_k$, the corresponding sequence of
solutions $\phi_{(i)}\in\cEc\V$ satisfies
\Beq
 \lim_{i\to\infty} \br\|\phi_{(i)}(t_0)\|_{\cH\CauV_k} =\infty.
\Eeq
\Erm

(4) The notion "completeness" has been chosen by analogy with the usual
completeness of flows (i. e. local one-parameter groups of automorphisms)
on manifolds. Indeed, any \sofe. determines a time evolution flow on the smf
$M\Cau$, and it is complete iff this flow is complete.

However, if making that rigorous, we have to circumvent the
difficulty that we are using a real-analytic calculus of superfunctions
while our flow is only differentiable in time direction.
Therefore, our flow is defined as an smf morphism $F: U\to M\Cau$ where
$\Space(U)\seq \Space(M\Cau)\times\mathbb R$ is open, but the topology and the
smf structure of $U$ are induced from the open embedding
$U\seq M\Cau\times\mathbb R_d$ where
$\mathbb R_d$ is the real axis equipped with the discrete topology and viewed
as a zero-dimensional smf. Explicitly,
one chooses a map $\theta(\cdot): \mathbb R_{>0}\to \mathbb R_{>0}$ such that
for any $c>0$, we have \eqref{ShortFrm} with $\theta:=\theta(c)$, and one sets
$\Space(U):=\bigcup_{c>0} cV\times(-\theta(c),\theta(c))$ where $V$ is the open
unit ball in $\Space(M\Cau)$. Now $F$ is given by
$\hat F|_{M\Cau\times\{t\}} = \Xi\sol[\Xi\Cau](t)$.

Analogously, every \sofe. determines flows on the smf's
$\L(\cH\CauV_k)$ of Cauchy data of Sobolev quality for all $k>d/2$; however,
as we saw in the previous remark, their completeness seems to be a stronger
condition than completeness of the \sofe.

\begin{proof}[Proof of the Theorem]
(ii)$\Rightarrow$(i) follows from Thm. \ref{UniqCor}.(ii).

(i)$\Rightarrow$(ii) follows from Prop. \ref{Lt4PEH} and Lemma \ref{SmCSols}.

(ii)$\Rightarrow$(iii):
For each $(t,x)\in\rdmm$ choose a buffer function $g\in\cD(\rdm)$
which is equal to one in some neighbourhood of $\J(\{(t,x)\})$,
and let $\phi(t,x):=\Xi\sol[g\phi\Cau](t,x)\allowbreak\in V$. It follows
from Thm. \ref{CausUniqThm} again that this does not depend on the
choice of $g$; hence $\phi:\rdmm\to V$ is well-defined. It also follows
directly from the construction that $\phi\in(\cE\V)\seven$ is the trajectory
wanted.

(iii)$\Rightarrow$(i): Given $\phi$ as in (i), it extends by (iii) to
a solution $\phi\in{\cE\V}\seven$. Let (say) $c=(a+b)/2$, and
choose $R$ with $\supp \phi(c)\seq \{x\in\rdm:\ \ \br|x|\le R\}$.
By Lemma \ref{SmCSols}, we get
$\supp \phi(t)\seq \{x\in\rdm:\ \ \br|x|\le R+\br|t-c|\}$; thus,
$\phi\in\cEc\V$, and the assertion becomes obvious.
\end{proof}

We conclude with a fairly simple sufficient criterion for
completeness.
First note that the definition \eqref{HkCauV}
of $\cH\CauV_k$ makes sense for all $k\ge0$; however, the assertions of Lemma
\ref{HklLemma} are valid only for $k>d/2$.

\begin{prp}
Suppose that a causal \sofe. satisfies the following additional conditions:

(i)  We have
\Beq
 \sup_{t\in(a,b)} \br\|\phi(t)\|_{\cH\CauV_0} < \infty
\Eeq
for every trajectory
$\phi\in C^\infty((a,b)\times \mathbb R^d) \otimes V\seven$
for which $\supp\phi(t)$ is compact for all $t\in(a,b)$.

(ii)
Let $k_0\in\mathbb Z$ be minimal such that $k_0+1>d/2$.
There exists a monotonously increasing function $F:\mathbb R_+\to \mathbb R_+$
such that
\Beq
 \br\|\partial_a\Delta[\phi\Cau|0]\|_{\cH\CauV_k} \le
 \br\|\phi\Cau\|_{\cH\CauV_{k+1}} F(\br\|\phi\Cau\|_{\cH\CauV_k})
\Eeq
for all $\phi\Cau\in(\cEc\CauV)\seven$,\ \
$a=1,\dots,d$,\ \ $k=0,\dots,k_0$.

Then the \sofe. is complete.
\end{prp}

\begin{proof}
Let $\phi$ be a trajectory as in (i).
We will prove inductively that \eqref{APEst} holds for for all
$k=0,\dots,k_0$, the start being given by (i). For the step,
we mimick the proof of Lemma \ref{SelfSm}: From the time-shifted
integral equation, we have for $0\le\theta <t$
\Beq
 \br\|\partial_a\phi\|_{\cH_k^{0,V}([\theta,t])}
 \le C_1 \|\partial_a\phi'(\theta)\|_{\cH_k\CauV}
 + C_2(b-\theta)\br\|\partial_a\Delta[\phi|0]\|_{\cH^{0,V}_k([\theta,t])};
\Eeq
investing the hypothesis (ii), the \rhs\ becomes
\Beq
 \le C_1 \|\partial_a\phi'(\theta)\|_{\cH_k\CauV}
 + C_2(b-\theta)
 (\br\|\phi\|_{\cH^{0,V}_k([\theta,t])}
 +\br\|\partial_a \phi\|_{\cH^{0,V}_k([\theta,t])} )
 F(\br\|\phi\|_{\cH^{0,V}_k([\theta,t])})
\Eeq
which again implies the assertion.
\end{proof}

\subsection{The main Theorem}
\begin{thm}\label{MainThm}
Suppose that the \sofe. is causal and complete.

(i) The formal solution $\Xi\sol$ is the Taylor expansion at zero of a
unique superfunctional
\Beqn SFuncXiSol
 \Xi\sol\in\M^{\cEc\V}(M\Cau),
\Eeq
and \eqref{SFuncXiSol} is an $M\Cau$-family of solutions of quality $\cEc$.

(ii) Consider the arising smf morphism (cf. Lemma \ref{CoordLem})
\Beq
 \check\Xi\sol: M\Cau\to M\cfg.
\Eeq
Its underlying map assigns to each bosonic Cauchy datum
$\CData$ the unique trajectory $\phi$ with $\phi(0)=\phi\Cau$.

(iii) The smf morphism $\alpha: M\cfg\to M\cfg$ given by
\Beqn Alpha
 \hat\alpha[\Xi]:=\Xi+ \Xi\sol[\Xi(0)]-\Xi\free[\Xi(0)]
\Eeq
is an automorphism of $M\cfg$ which makes the diagram
\begin{eqnarray*}
 &M\Cau & \\
 \check\Xi\free&\swarrow\hskip1cm\searrow&\check\Xi\sol \\
 M\cfg & \too[\qquad\alpha\qquad]{} &M\cfg\\
\end{eqnarray*}
commutative.

(iv) The image of the $\check\Xi\sol$ is a split sub-smf which we call the
{\em smf of classical solutions}, or, more exactly, the
{\em smf of smooth classical solutions with spatially compact support},
and denote by $M\sol\seq M\cfg$.

(v) $M\sol$ has the following universal property:
Recall that fixing an smf $Z$ we have a bijection between $Z$-families
$\Xi'$ of configurations of quality $\cEc$ with time definition interval
$\mathbb R$, and morphisms $\check\Xi':Z\to M\cfg$. Now $\Xi'$ is a solution
family iff $\check\Xi'$ factors to $\check\Xi':Z\to M\sol\seq M\cfg$.

In this way, we get a bijection between $Z$-families
$\Xi'$ of solutions of quality $\cEc$ with time definition interval
$\mathbb R$, and morphisms $\check\Xi':Z\to M\sol$.

(vi) The underlying manifold $\widetilde{M\sol}$ identifies with the set of all
bosonic configurations which satisfy the field equations with all
fermion fields put to zero:
\Beq
 \widetilde{M\sol} = \{\phi\in (\cEc\V)\seven:\quad
 L_i[\phi|0]=0,\ \ i=1,\dots,N\}.
\Eeq

\end{thm}

\Brm
(1) In the language of category theory, assertion (v) means that the
cofunctor
\Beq
 \{\text{supermanifolds}\}\to\{\text{sets}\},\qquad
 Z\mapsto\{\text{$Z$-families of solutions}\}
\Eeq
is represented by the object $M\sol$ with the universal element $\Xi\sol$.
Thus, \eqref{SFuncXiSol} is the {\em universal family of solutions of quality
$\cEc$}: Every other family of solutions of this quality arises
uniquely as pullback from \eqref{SFuncXiSol}.

(2) Note that $M\sol$ is still a linear smf which is, however, in a non-linear
way embedded into $M\cfg$. In the successor paper we will show that,
once the action of the Lorentz group on $V$ has been fixed, the
sub-smf $M\sol$ is invariant under the arising action of the Poincar\'e group
on $M\cfg$; the other data $\alpha$, $\check\Xi\free$, $\check\Xi\sol$ are not
(they are only invariant under the Euclidian group of $\rdm$).

(3) The superfunction \eqref{SFuncXiSol} is uniquely characterized by the
conditions \eqref{XiSolIsSol}. The initial value condition
can be recoded supergeometrically to the fact that the composite morphism
\Beq
  M\Cau \too{\check\Xi\sol} M\cfg \too\pi M\Cau
\Eeq
is the identity; here $\pi$ is the projection onto the Cauchy data:
\Beqn DefPi
 \hat\pi[\Xi] = \Xi(0).
\Eeq

(4) The Taylor expansion of \eqref{Alpha} at zero is just
\Beq
 \hat\alpha[\Xi]_{(0,0)}=\Xi+\Xi\sol_{(\ge3)}[\Xi(0)].
\Eeq

(5) Although a Inverse Function Theorem is valid for morphisms of smf's
with Banach model space (it will be given in the next part of
\cite{[IS]}), we cannot apply it to $\alpha$ since the model space $\cEc$
is far away from being Banach. Fortunately, a purely soft method well-known
in category theory will turn out to be sufficient to conclude that
$\alpha$ is even globally isomorphic.

(6) In qft slang, the homomorphism
\Beqn RestMSh
 \O^F(M\cfg)\to\O^F(M\sol),\qquad K[\Xi]\mapsto K[\Xi\sol]
\Eeq
is called {\em restriction of classical observables onto the mass shell}
(the latter term comes from free field theory). It follows from ass. (iii) and
the proof given below that \eqref{RestMSh} is surjective.

(7) Let us comment on the fact that completeness depends only on the
underlying bosonic \sofe.: Mathematically, this is an analogon of several
theorems in supergeometry that differential-geometric
tasks, like trivializing a fibre bundle, or presenting a closed form
as differential, are solvable iff the underlying smooth tasks are solvable.

Physically, our interpretation is somewhat speculative:
In the bosonic sector, the classical field theory approximates the behaviour
of coherent states, and completeness excludes that "too many" particles may
eventually assemble at a space-time point, making the state non-normable.
On the fermionic side, apart from the non-existence of genuine coherent
states, it is the Pauli principle which automatically prevents such an
assembly.
\Erm

\begin{proof}[Proof of the main Theorem]

Ad (i), (ii). Apply Thm. \ref{cEcFit} with $Z:=M\Cau$ and
$\Xi\CauP=\CData$.

Ad (iii). The commutativity of the diagram is clear; we show that
$\alpha$ is isomorphic. Let
\Beq
  \cEc\free := \{\xi\in\cEc\V:\quad (\partial_t+ K(\partial_x))\xi =0\},\quad
  \cEc\zero := \{\xi\in\cEc\V:\quad \xi(0)=0\}.
\Eeq
Using Lemma \ref{CauUniq}.(iii), we have a direct decomposition
\Beqn DecCEc
  \cEc\V=\cEc\free\oplus\cEc\zero,
\Eeq
with the corresponding projections given by
\Beq
 \pr\free(\xi):=\Xi\free[\xi(0)],\quad \pr\zero:=1-\pr\free
\Eeq
(we recall that due to linearity, the insertion makes sense, as in
Cor. \ref{EstXiFree}).

Now \eqref{DecCEc} yields an identification
$M\cfg = \L(\cEc\V) = \L(\cEc\free) \times \L(\cEc\zero)$,
with the corresponding projection morphisms being
$\L(\pr\free)$, $\L(\pr\zero)$, and $\alpha$ becomes the composite
\Beqn AlphIsComp
 M\cfg = \L(\cEc\free) \times \L(\cEc\zero)
  \too{(\check\Xi\sol\circ\pi)\times\L(\seq)}
  M\cfg\times M\cfg \too{\L(+)} M\cfg
\Eeq
with $\pi$ as in \eqref{DefPi}. As often in supergeometry, it is
convenient to look at the point functor picture, i.~e. we look how
$\alpha$ acts on $Z$-families of configurations: For any smf $Z$ we get a map
\Beqn Yoneda
 \opn Mor (Z,M\cfg) \to \opn Mor (Z,M\cfg),\quad \xi\mapsto\alpha\circ\xi,
\Eeq
and our assertion follows once we have shown that this is always
an isomorphism. (Indeed, it is sufficient to take $Z=M\cfg$, $\xi=\opn Id $.)

Now $\alpha$ acts on $\xi\in\M^{\cEc}(Z) = \opn Mor (Z,M\cfg)$ by
\Beq
 \xi = \xi\free + \xi\zero\mapsto \Xi\sol[\xi\free(0)] + \xi\zero.
\Eeq
We show injectivity of \eqref{Yoneda}: If
$\alpha\circ\xi=\alpha\circ\xi'$
then, taking Cauchy data at both sides, we get that $\xi\free$,
$(\xi')\free$ have the same Cauchy data;
hence $\xi\free=(\xi')\free$, and the hypothesis now implies $\xi=\xi'$.

We show surjectivity of \eqref{Yoneda}: Given $\xi\in\M^{\cEc}(Z)$, its
preimage is given by $\xi\free + \xi\zero$ with
\Beq
 \xi\zero:=\xi - \Xi\sol[\xi(0)],\quad \xi\free:= \Xi\free[\xi(0)].
\Eeq
Assertion (iii) is proved.

Assertion (iv) follows from (iii), and (vi) follows from (ii)
(or by considering $P$-families where $P$ is a point).

Finally, (v) follows from:

\begin{prp}
A $Z$-family $\Xi'$ of configurations of quality $\cEc$ with time definition
interval $\mathbb R$ is a solution family iff the corresponding morphism
$\check\Xi'$ coincides with the composite
\Beq
 Z \too{\check\Xi'} M\cfg \too\pi M\Cau \too{\check\Xi\sol} M\cfg.
\Eeq
Here $\pi$ is the projection onto Cauchy data, $\hat\pi = \Xi(0)$.
\end{prp}

\begin{proof}
"$\Leftarrow$" follows since $\Xi\sol$ is a solution family while
"$\Rightarrow$" follows using Thm. \ref{UniqCor}.
\end{proof}

The Theorem is proved.
\end{proof}

{}From Thm. \ref{CausUniqThm}.(ii) we get a "supergeometric" formulation of
causality:

\begin{prp}
Let be given an smf $Z$, two morphisms $Z\rightrightarrows M\Cau$, and
an open set $U\seq\rdm$ such that the composites
\Beq
 Z\rightrightarrows M\Cau \too{\opn proj }
  \L(C^\infty(U)\otimes V)
\Eeq
coincide. Let $U':=\{(t,x)\in\rdmm: \J(\{(t,x)\})\seq U\}$. Then the
composites
\Beq
 Z\rightrightarrows M\Cau \too{\check\Xi\sol} M\cfg
  \too{\opn proj } \L(C^\infty(U')\otimes V)
\Eeq
coincide.
\qed\end{prp}

\subsection{Variation: The space $\protect{\cEt}$}\label{SpaceCit}

There is a slightly larger space than $\cEc$ such that $\cD(\rdm)$
still appears as space of corresponding Cauchy data.

Let $\cEt$ be the space of all $f\in C^\infty(\rdmm)$ such
that for every Poincar\'e transformation $l:\rdmm\to\rdmm$ and every
time $\theta>0$, $\supp l^*(f) \cap [-\theta,\theta]\times\rdm$ is compact.
It is sufficient to check this for Lorentz transformations $l$ only.
We will equip $\cEt$ with the topology defined by the seminorms
\Beq
  \br\|f\|_{l,\theta,w,k}:= \sup_{x\in\rdm,\ \br|t|\le\theta}
  w(x) \sum_{\br|\nu|\le k} |\partial^\nu l^*(f)(t,x)|
\Eeq
where $l$ is a Lorentz transformation, $\theta>0$, $k>0$, and $w$ is a
non-negative continuous function on $\rdm$.

We have a continuous inclusion $\cEc\seq\cEt$ which is proper for $d\ge1$.
For instance, for $f(t,x) := \sum_{n\in\mathbb Z_+} \varphi(t-n,x-2^np)$ where
$p\in\rdm\setminus0$, and $\varphi\in\cD(\rdmm)$, the intersection of $\supp f$
with any space-like hyperplane in $\rdmm$ is compact; hence
$f\in\cEt\setminus\cEc$.

We get a morphism of the corresponding configuration smf's
\Beq
 M\cfg \too{\L(\seq)} M_{\opn t } := \L(\cEt\V)
\Eeq
where $\cEt\V:=\cEt\otimes V$. The map
$\pi_{\opn t } : \cEt\V \to \cEt\CauV$
(assignment of Cauchy data) is still well-defined, continuous, and
surjective; thus, the morphism of assignment of Cauchy data,
$M\cfg \too{\L(\pi)} M\Cau$, factors to
$M\cfg \too{\L(\seq)} M_{\opn t } \too{\L(\pi_{\opn t })} M\Cau$.

\begin{prp}
Suppose that the \sofe. is causal and complete. For any smf $Z$, the map
\Beq
 \M^{\cEc\V}(Z) \too\seq \M^{\cEt\V}(Z),
\Eeq
maps the solution families of quality $\cEc$ bijectively onto the
solution families of quality $\cEt$. Thus, the composite embedding
$M\sol \seq M\cfg \too{\L(\seq)} M_{\opn t }$ makes $M\sol$ the
solution smf for the quality $\cEt$.
\end{prp}

\begin{proof}
Given a solution family $\Xi'\in\M^{\cEt\V}(Z)$, its Cauchy data
$\pi_{\opn t } \Xi'\in\M^{\cEc\CauV}(Z)$ determine a solution family
$\Xi":=\Xi\sol[\pi_{\opn t }\Xi']\in\M^{\cEc\V}(Z)$;
by Thm. \ref{UniqCor}.(ii), we have $\Xi'=\Xi"$.
\end{proof}

\subsection{The smf of classical solutions without support restriction}

It takes not much additional effort to lift the constraints on the
supports of solution families, considering arbitrary smooth solution families.
The appropiate smf's of Cauchy data and configurations are
\Beq
 M\Cau_{C^\infty} := \L(\cE\CauV),\quad M\cfg_{C^\infty} := \L(\cE\V).
\Eeq

\begin{thm}\label{MainThmSm}
Suppose the \sofe. is causal and complete.

The formal solution $\Xi\sol$ is the Taylor series at zero of
a unique superfunctional $\Xi\sol\in\M^{\cE\V}(\allowbreak M\Cau_{C^\infty})$,
which in turn determines an smf morphism
\Beq
 \check\Xi\sol:M\Cau_{C^\infty} = \L(\cE\CauV)\to\L(\cE\V) = M\cfg_{C^\infty}.
\Eeq
Its underlying map assigns to each bosonic Cauchy datum
$\CData$ the unique trajectory $\phi$ with $\phi(0)=\phi\Cau$.

\eqref{Alpha} defines an smf automorphism
$\alpha: M\cfg_{C^\infty}\to M\cfg_{C^\infty}$
which satisfies $\alpha\circ\check\Xi\free=\check\Xi\sol$ again.

The image of $\check\Xi\sol$ is a split sub-smf which we call the
{\em smf of smooth classical solutions without support restriction},
and denote by $M\sol_{C^\infty}\seq M\cfg_{C^\infty}$.

A $Z$-family $\Xi'$ of quality $\cE$
is a solution family iff the corresponding morphism
$\check\Xi':Z\to M\cfg_{C^\infty}$
factors to $\check\Xi':Z\to M\sol_{C^\infty}\seq M\cfg_{C^\infty}$.

In this way, we get a bijection between $Z$-families
$\Xi'$ of solutions of quality $\cE$,
and morphisms $\check\Xi':Z\to M\sol_{C^\infty}$.
\end{thm}

\begin{proof}
The proof is quite analogous to that of Thm. \ref{MainThm};
Lemma \ref{XiCDWR} provides the needed Taylor expansions, and
Thm. \ref{cEcFitWR} shows that they fit together.
\end{proof}

We get a commutative diagram of smf's
\Bcd
 M\Cau           @>\check\Xi\sol>> M\sol           @>\seq>> M\cfg\nt
 @V\seq VV                         @V\seq VV                @V\seq VV\nt
 M\Cau_{C^\infty}@>\check\Xi\sol>> M\sol_{C^\infty}@>\seq>> M\cfg_{C^\infty}.
\Ecd

\subsection{Local excitations}\label{LocExc}
A further variant arises by considering compactly supported excitations of
a classical solution; in particular, it is applicable for situations with
spontaneous symmetry breaking, like the Higgs mechanism:

\begin{thm}
Suppose that the \sofe. is causal and complete, and fix a trajectory
$\phi\in {\cE\V}\seven$; let $\CData$ be its Cauchy data.

(i) The superfunctional
\Beqn DefXiExc
 \Xi\exc_\phi[\Xi\Cau] := \Xi\sol[\Xi\Cau+\phi\Cau]-\phi,
\Eeq
which lies a priori in $\M^{\cE\V}(M\Cau_{C^\infty})$, restricts to a
superfunctional
\Beq
 \Xi\exc_\phi[\Xi\Cau]\in\M^{\cEc\V}(M\Cau).
\Eeq

(ii) Consider the arising smf morphism
$\check\Xi\exc_\phi: M\Cau \to M\cfg$.
The smf morphism $\alpha_\phi: M\cfg\to M\cfg$ given by
\Beq
 \widehat{\alpha_\phi}[\Xi]:=
 \Xi+ \Xi\sol[\Xi(0)+\phi\Cau]-\Xi\free[\Xi(0)] - \phi
\Eeq
is an automorphism of $M\cfg$ which satisfies
$\alpha_\phi\circ\check\Xi\free=\check\Xi\exc_\phi$.

(iii) The image of $\check\Xi\exc_\phi$ is a split sub-smf
which we call the {\em smf of excitations around the trajectory $\phi$},
and denote by $M\exc_\phi\seq M\cfg$.

(iv) $M\exc_\phi$ has the following universal property:
Given a $Z$-family $\Xi'\in\M^{\cEc\V}(Z)$, the corresponding morphism
$\check\Xi':Z\to M\cfg$ factors through $M\exc_\phi$ iff
the $Z$-family $\Xi'+\phi\in\M^{\cE\V}(Z)$ is a solution family.
\end{thm}

\begin{proof}
Ad (i). First, we prove that
for $\phi\CauP\in(\cEc\CauV)\seven$, we have
\Beqn ExcTaySer
 \Xi\exc_\phi[\Xi\Cau]_{\phi\CauP} \in \P(\cEc\CauV;\cEc\V).
\Eeq
Indeed,
\Beq
 \Xi\exc_\phi[\Xi\Cau]_{\phi\CauP} =
 \Xi\sol_{\phi\Cau+\phi\CauP}[\Xi\Cau]- \phi.
\Eeq
On the other hand, setting
\Beq
 \phi":=\Xi\sol[\phi\Cau+\phi\CauP|0],
\Eeq
it follows from Thm. \ref{CausUniqThm} that $\phi"-\phi\in(\cEc\V)\seven$.
Now \eqref{ExcTaySer} follows from Prop. \ref{ExcAnal}.

Now, we use again the strictly separating set (cf.
\whatref{\StrictSep}) of linear functionals
\eqref{StrSepSet}. Since all these functionals extend onto $\cE\V$, it follows
simply from Thm. \ref{MainThmSm} that the elements \eqref{ExcTaySer} fit
together to the superfunction wanted.

The proofs of the remaining assertions are quite analogous to those
for Thm. \ref{MainThm}; in \eqref{AlphIsComp} and the following formulas, one
simply replaces $\Xi\sol$ by $\Xi\exc_\phi$.
\end{proof}

\Brm
This theorem yields new information only if the $\CData$ are not compactly
carried. If they are, i.~e. $\CData\in(\cEc\CauV)\seven$, then
\eqref{DefXiExc} is already a priori defined as element of
$\M^{\cEc\V}(M\Cau)$, and $M\exc_\phi$ can be identified with $M\sol$.
\Erm

\subsection{Other generalizations}
For a non-causal \sofe., one can still construct solution supermanifolds
of Sobolev quality. Adapting the proof of
Thm. \ref{MainThm} we have:

\begin{cor}
Let be given a (perhaps non-causal) \sofe., and fix $k,l$ with $k>d/2+\mu l$.
Suppose that
\Beqn APEstAnew
 \sup_{t\in(a,b)} \br\|\phi(t)\|_{\cH\CauV_k} < \infty
\Eeq
holds for every trajectory $\phi\in\cH_k^{l,V}(I)\seven$.
Then the formal solution is the
Taylor expansion at zero of a uniquely determined superfunction
$\Xi\sol[\Xi\Cau]\in\M^{\cH_k^{l,V}(\mathbb R)}(\L(\allowbreak\cH_k\CauV))$
which is
the universal solution family for the quality $\cH_k^l$.
The image of the corresponding smf morphism
$\check\Xi\sol:\L(\cH_k\CauV)\to\cH_k^{l,V}(\mathbb R))$ is a
sub-supermanifold which is called the
{\em supermanifold of solutions of quality $\cH_k^l$}.
\qed\end{cor}

One can get rid of the $k$-dependence by taking the intersections over all $k$
equipped with the projective limes topology:
\Beq
 \cH_\infty\CauV := \bigcap_{k>0} \cH_k\CauV,\qquad
 \cH_\infty\V(\mathbb R):= \bigcap_{k,l>0} \cH_k^{l,V}(\mathbb R).
\Eeq
One then gets a morphism
\Beq
 \check\Xi\sol:\L(\cH_\infty\CauV)\to\L(\cH_\infty\V(\mathbb R))
\Eeq
provided \eqref{APEstAnew} holds for every trajectory
$\phi\in\cH_\infty\V(\mathbb R)\seven$.

Note that, roughly spoken, $\cH_\infty\CauV$, $\cH_\infty\V(\mathbb R)$
"lie between" the Schwartz spaces ${\mathcal S}$ and $C^\infty$;
even for a causal \sofe., it is not clear whether one can descend to the
Schwartz spaces.

Let us sketch an abstract version of our approach: we start with a
$\mathbb Z_2$-graded Banach space $B$ and a strongly continuous
group $(\A_t)_{t\in\mathbb R}$ of parity preserving operators;
let $K: \opn dom K\to B$ denote the
generator of this group. Also, let be given an entire superfunction
$\Delta=\Delta[\Xi]\in \M^B(\L(B))$ the Taylor expansion $\Delta_0$ of which
in zero has lower degree $\ge2$ and satisfies $\Delta_0\in\P(B,cU;B)$ for all
$c>0$ where $U$ is the unit ball of $B$.
Formally, the equation of interest is
\Beqn AbstrDEq
 \frac d{dt} \Xi'=K\Xi' + \Delta[\Xi'];
\Eeq
however, this makes sense only if $\Xi'$ takes values in $\opn dom K$.
Therefore we look for the integrated version
\Beqn AbstrIEq
 \Xi'(t) = \A_t\Xi'(0) + \int ds f(t,s)\A_{t-s}\Delta[\Xi'](s)
\Eeq
where $f(t,s)$ is as in \eqref{DefFTS}.

For a connected subset $I\seq \mathbb R$, $I\owns0$  with non-empty kernel, let
$B(I):=C(I,B)$ equipped with the topology induced by the seminorms
$\br\|\phi\|_{B([a,b])}:= \max_{t\in[a,b]} \br\|\phi(t)\|$
where $a,b\in I$, $a<b$. If $Z$ is any smf we call an element
$\Xi'\in\M^{B(I)}(\L(B))$ a {\em solution family} iff \eqref{AbstrIEq} holds.
(Of course, if $\Xi'$ takes values in $D:=\opn dom K$ (i.~e.
$\Xi'\in\M^{C(I,\opn dom K)}(\L(B))$ where $\opn dom K$ is equipped with
the graph norm) then \eqref{AbstrIEq} is equivalent to the differentiated
form \eqref{AbstrDEq}).

Now our approach generalizes to:

\begin{cor}

(i) There exists a formal solution
$\Xi\sol=\Xi\sol[\Xi\Cau]\in\P(B;B(\allowbreak\mathbb R))$
of \eqref{AbstrIEq}, with $\Xi\sol_{(\le1)}(t)=\A_t\Xi\Cau$, and
\Beq
 \Xi\sol_{(n)}(t) = \int ds f(t,s)\A_{t-s}\Delta[\Xi\sol_{(\le n-1)}]_{(n)}(s)
\Eeq
for $n\ge2$.

(ii) For any $c>0$ there exists $\theta>0$ such that
$\Xi\sol\in\P(B,cU;B([-\theta,\theta]))$ where $U\seq B$ is the unit ball.

(iii) Suppose that for any solution family $\Xi'\in\M^{B(I)}(\L(B))$, we have
$\sup_{t\in I} \|\phi(t \allowbreak)\| \allowbreak < \allowbreak\infty$.
Then the formal solution is the
Taylor expansion at zero of a uniquely determined superfunction
$\Xi\sol[\Xi\Cau]\in\M^{B(\mathbb R)}(\L(B))$ which is the universal
solution family. The image of the corresponding smf morphism
$\check\Xi\sol:\L(B)\to\L(B(\mathbb R))$ is a sub-supermanifold which is called
the {\em solution supermanifold} of \eqref{AbstrIEq} or \eqref{AbstrDEq}.
\qed\end{cor}

\subsection{Solutions with values in Grassmann algebras}
\label{SolValGrass}
The most naive notion of a configuration in a classical field model with
anticommuting fields arises by replacing the domain $\mathbb R$ for the real
field components by a finite-dimensional Grassmann algebra
$\Lambda_n=\mathbb C[\zeta_1,\dots,\zeta_n]$ (we recall that, in accordance
with
our hermitian framework, only complex Grassmann algebras should be used).
Here we consider only smooth
configurations; thus, a {\em $\Lambda_n$-valued configuration}
is a tuple $\xi=(\phi|\psi)$ with
\Bea
 &\phi_i\in C^\infty(\rdmm,(\Lambda_n)\sevR)
  &\ \text{for $i=1,\dots,N\seven$},\\
 &\psi_j\in C^\infty(\rdmm,(\Lambda_n)\sodR)
  &\ \text{for $j=1,\dots,N\sodd$}.
\Eea
Now, comparing with \whatref{\FamPhilo} we see that
$\xi$ encodes just a $Z_n$-family over $\mathbb R$ of quality
$\cE$ where $Z_n$ is the $0|n$-dimensional smf, so
that $\O(Z_n)=\Lambda_n$. Also, $\xi$ is a solution family in our
sense iff the field equations are satisfied in the plain sense. We now
get an overview over all $\Lambda_n$-valued solutions:

\begin{cor}\label{GrassmSols}
Suppose that the \sofe. is causal and complete, and let be given
$\Lambda_n$-valued Cauchy data
\Beq
 \xi\Cau\in C^\infty(\rdm, (\Lambda_n\ \otimes\ V)\sevR).
\Eeq
Then there exists a unique $\Lambda_n$-valued solution $\xi=(\phi|\psi)$
with these Cauchy data. It is given by
\Beq
 \xi = \left(\check\Xi\sol\circ(\xi\Cau)\spcheck\right)\sphat
 =\Xi\sol_{b(\phi\Cau)}[s(\xi\Cau)]
\Eeq
where $b(\cdot): \Lambda_n\to\mathbb C$ denotes the body map,
and $s(\cdot)= 1-b(\cdot)$ the soul map.
\qed\end{cor}

We now look for solutions in the infinite-dimensional Grassmann algebra
$\Lambda_\infty$ of supernumbers introduced by deWitt \cite{[DeWitt]}:
\Beqn LambdaDirLim
 \Lambda_\infty = \bigcup_{n>0} \Lambda_n = \lim_{\longrightarrow} \Lambda_n .
\Eeq
Let $\mathbb R^\infty$ be the vector space of all number sequences
$(a_i)_{i\ge1}$, equipped with the product topology. The topological dual,
$(\mathbb R^\infty)^*$, is algebraically generated by the projections
$\pi_i$ on the $i$-th component.

Let $Z_\infty:=\L(\Pi \mathbb R^\infty)$; this is an infinite-dimensional
smf the underlying manifold of which is a single point.
(We recall that $\Pi$ is just an odd formal symbol.)
Now the elements $\zeta_i:= e_i\circ\Pi^{-1}$ lie in the odd part of
$(\Pi \mathbb R^\infty)^*\seq \O(Z_\infty)$; from the universal
property of the Grassmann algebra we get an algebra homomorphism
\Beq
 \Lambda_\infty \to \O(Z_\infty),
\Eeq
and one shows that this is an isomorphism; thus, we can identify both sides.

Now, fixing some $k\ge0$, and given an element
$f\in\O^{C^\infty(\mathbb R^k)}(Z_\infty)$, we get a map
\Beqn FPrime
 f':\mathbb R^k\to \Lambda_\infty,\quad x\mapsto \delta_x\circ f
\Eeq
which has the property that for any bounded open $U\seq \mathbb R^k$ coincides
with a $C^\infty$ map $f'|_U: U\to \Lambda_n$ for sufficiently large $n$.
In this way, we get an isomorphism
\Beq
 \O^{C^\infty(\mathbb R^k)}(Z_\infty) \too\cong
 C^\infty(\mathbb R^k, \Lambda_\infty)
\Eeq
where we equip $\Lambda_\infty$ with the locally convex inductive limit
topology arising from \eqref{LambdaDirLim}.

Thus, a $\Lambda_\infty$-valued smooth configuration
$\xi\in (C^\infty(\rdmm, \Lambda_\infty)\otimes V)\seven$ encodes the same
information as an smf morphism $Z_\infty\to M\cfg$, i.~e. a $Z_\infty$-valued
point of $M\cfg$; analogously for the Cauchy data.

It follows that Cor. \ref{GrassmSols} holds also for $n=\infty$.

\Brm
An element $f\in\O^{\cE}(Z_\infty)$ lies in $\O^{\cEc}(Z_\infty)$ iff
for each sequence $i_1<\dots<i_n$ of indices, the function
$(\partial_{\zeta_{i_1}}\cdots\partial_{\zeta_{i_n}} f)\sptilde\in
C^\infty(\rdmm)$ has its support in some $\bV_r$. The function
\eqref{FPrime} itself needs not to have this support property.
\Erm

\subsection{Examples}
Here we consider only two characteristic examples;
a systematic exploration of a large class of classical field theories will
be given in the successor paper.

\subsubsection{The $\Phi^4$ model}
Let us see how this model, considered in \ref{Phi4}, fits into our model
class: Here
$(d,N\seven|N\sodd,\sd,(L_i[\ul\Xi]))= (3,2|0,(1,0),(L_i[\ul\Phi]))$
where $L_1,L_2$ are given by \eqref{L1L2}; thus
$K = \begin{pmatrix} 0 & -1 \\ K' &0 \end{pmatrix}$
with $K'=-\sum^3_{a=1}{\partial_a}^2$ and
$\Delta=(0,4q\Phi_1^3)$. Now
\Beq
 A = \begin{pmatrix}
  \partial_t\cA & \cA \\
  {\partial_t}^2\cA  & \partial_t\cA
 \end{pmatrix},
\Eeq
where $\cA$ is given by \eqref{PJFunc}, and the conditions \eqref{TheSmCEst},
\eqref{SmCond} as well as the causality condition \eqref{TheCausEsts}
are easily checked. It is a classical result that this \sofe. is complete.
Since the kinetic operator is spatially second order but $\cA$ improves
spatial smoothness by one degree, the "smoothness loss" $\smloss$ is one.
(The smoothness condition would allow even a derivative coupling;
however, such an interaction could not stem from a Lagrangian.)

\subsubsection{The Thirring model}
This is purely fermionic, with $d=1$ and field contents
given by a fermionic Dirac spinor field $\Psic=(\Psic_1,\Psic_2)\trp$;
thus, the real field components are
$(\Psi_i)^4_{i=1}=(\opn Re \Psic_1,\opn Im \Psic_1,\allowbreak
 \opn Re \Psic_2,\allowbreak \opn Im \Psic_2)$.
For $a=0,1$, set
\Beq
 j^a(\ulPsic):=\dcj{\ulPsic}\gamma^a\ulPsic;
\Eeq
where $\gamma^0,\gamma^1$ are the usual gamma matrices, and
$\psi\mapsto \dcj{\psi}:=\cj{\psi}\trp\gamma^0$ denotes Dirac
conjugation. We consider the Thirring model in its classical form,
with Lagrangian
\Beq
 {\mathcal L}[\ulPsic] = \frac \i2 \sum_{a=0}^1 \br(
  \dcj{\ulPsic}\gamma^a\partial_a\ulPsic -
  \partial_a\dcj{\ulPsic}\gamma^a\ulPsic) -
  \frac g2 \sum_{a=0}^1 j_a(\ulPsic)j^a(\ulPsic)
\Eeq
where $g\in\mathbb R$.

The variational derivatives are
\Beq
 \frac \delta{\delta\dcj{\ulPsic}_\alpha}{\mathcal L}[\ulPsic]
 = \bigl(\sum_a
 \gamma^a(\i\partial_a\ulPsic - g j_a(\ulPsic)\ulPsic)\bigr)_\alpha,
 \quad \alpha=1,2,
\Eeq
and its hermitian conjugate; thus, the "field equations" are equivalent
to the vanishing of
\Beq
 L^{\opn c }[\ul\Psi]:= \partial_t\ulPsic +
 (\gamma^0)^{-1}\gamma^1\partial_1\ulPsic
 + \i g (\gamma^0)^{-1} \sum_a \gamma^a j_a(\ulPsic)\ulPsic.
\Eeq
It is easy to check that we get a causal \sofe.
\Beq
(d,N\seven|N\sodd,\sd,(L_i[\ul\Xi])) :=
(1,0|4,(0,0,0,0),
 (\opn Re L^{\opn c }_1[\ul\Psi],\opn Im L^{\opn c }_1[\ul\Psi],
  \opn Re L^{\opn c }_2[\ul\Psi],\opn Im L^{\opn c }_2[\ul\Psi])).
\Eeq
Since the underlying bosonic \sofe. is "empty", this
model (as well as its generalization, the Gross-Neveu model), is complete.
It is well known that this model can be explicitly solved; using the
method given e.~g. in \cite[11.2.A]{[BigBog]}
one gets a closed formula for $\Psi\sol$.
We use complex notations, setting
\Beq
 \Psi\cCau:= (\Psi_1\Cau + \i\Psi_2\Cau,\ \Psi_3\Cau + \i\Psi_4\Cau)\trp
 \in\O^{\cEc\Cau\otimes\mathbb C^2}(M\Cau)
\Eeq
and analogously for $\Psi\csol\in\O^{\cEc\otimes\mathbb C^2}(M\Cau)$. Also, set
\Beq
 \phi_2\Cau := -\dcj{\Psi\cCau}\gamma^0\Psi\cCau,\quad
 \phi_1\Cau(x) :=  \int_0^x dx'\dcj{\Psi\cCau}(x')\gamma^1\Psi\cCau(x').
\Eeq
Let $\Phi\free_1[\Phi\Cau_1,\Phi\Cau_2]$ be
the solution operator of the massless free scalar field:
\Beq
 \Phi\free_1[\Phi\Cau_1,\Phi\Cau_2](t,y) =
 \int_{\mathbb R^3} dx
  \left(\partial_t\cA(t,y-x)\Phi\Cau_1(x)
  + \cA(t,y-x)\Phi\Cau_2(x)\right),
\Eeq
where $A(t,x)$ is the massless Pauli-Jordan exchange function given by its
spatial Fourier transform as
$\hat\cA(t,p) = (2\pi p^2)^{-1/2}\sin (\sqrt{p^2}t)$.
Now the solution operator for the free Dirac field is given by
\Beq
 \Psi\cfree[\Phi\cCau](t,y) =
 \int_{\mathbb R^3} dx
 \sum_{a=0}^1 (\gamma^a\partial_a \cA)(t,y-x)\Phi\cCau(x).
\Eeq

\begin{prp}
We have
\Beq 
  \Psi\csol[\Psi\cCau] = \exp(\i g\Phi_1\free[\phi\Cau_1,\phi\Cau_2])\;
  \Psi\cfree[\exp(-\i g\phi\Cau_1)\Psi\cCau].
\Eeq
\qed\end{prp}

All the smf's $M\Cau,M\cfg, M\sol$ connected with this model have a
single point $P$ as underlying manifold; the whole information of
$\check\Xi\sol: M\Cau\to M\cfg$ lies in the homomorphism
$\check\Xi\sol: \O_{M\cfg}(P)\to\O_{M\Cau}(P)$ of infinite-dimensional
Grassmann algebras.

It would be interesting to see how $M\sol$ is connected with the tree
approximation of the quantized model.

\appendix
\section{Table of spaces of configurations and Cauchy data}

\begin{tabular}{l||l|l|l}
 Quality
 & smooth ($C^\infty$)
 &$C^\infty$
 &Sobolev
\\
 &
 &with causal support
 &($k>\frac d2$)
\\ \hline
 Cauchy data
 &$\cE\Cau$
 &$\cEc\Cau$
 &$\cH_k\Cau$
\\
 for components
 & $=C^\infty(\rdm)$
 & $=\cD(\rdm)$
 & $= H_k(\rdm)$
\\ \hline
 Cauchy data
 &$\cE\CauV$
 &$\cEc\CauV$
 &$\cH_k\CauV$
\\
 &$=C^\infty(\rdm,V)$
 &$=\cD(\rdm,V)$
 &(cf. \eqref{HkCauV})
\\ \hline
 Configurations
 &$\cE$
 &$\cEc$
 &$\cH_k^l(I)$
\\
 for components
 &$=C^\infty(\rdmm)$
 &(cf. \ref{SpaceCic})
 &(cf. \eqref{Hkl})
\\ \hline
 Configurations
 & $\cE\V$
 & $\cEc\V$
 & $\cH_k^{l,V}(I)$
\\
 & $=C^\infty(\rdmm,V)$
 & $=\cEc\otimes V$
 & (cf. \eqref{HkVl})
\\ \hline
 Use
 & Variant of
 & Main Thm.
 & Technical
\\
 & Main Thm.
 &
 &
\end{tabular}

\vfill

{\sc Technische Universit\"at Berlin}

Fachbereich Mathematik, MA 7 -- 2

Stra\ss e des 17. Juni 136

10623 Berlin

FR Germany

\medskip

{\em E-Mail  address: } schmitt@math.tu-berlin.de

\begin{thebibliography}{99}
\bibitem{[BigBog]}
 Bogoliubov N N, Logunov A A, Oksak A I, Todorov I T:
 General principles of quantum field theory (in russian).
 "Nauka", Moskwa 1987
\bibitem{[ChoYM]}
Choquet-Bruhat Y, Christodoulou D:
 Existence of global solutions of the Yang-Mills, Higgs and spinor field
 equations in $3+1$ dimensions.
 Ann. de l'E.N.S., $4^{\text{\`eme}}$ s\'erie, Tome 14 (1981), p. 481--500
\bibitem{[ChoSugr]}
Choquet-Bruhat Y:
 Classical supergravity with Weyl spinors.
 Proc. Einstein Found. Intern. Vol. 1, No. 1 (1983) 43-53
\bibitem{[DeWitt]}
DeWitt B:
 Supermanifolds.
 Cambridge University Press, Cambridge 1984
\bibitem{[Eardley/Moncrief]}
Eardley D M, Moncrief V:
 The global existence of Yang-Mills-Higgs Fields in 4-dimensional
 Minkowski space.
 I. Local Existence and Smoothness Properties.
 II. Completion of the proof.
 Comm. Math. Phys. 83, 171--191 and 193--212 (1982)
\bibitem{[Ginebre/Velo]}
 Ginibre J, Velo G: The Cauchy Problem for coupled Yang-Mills and
 scalar fields in the temporal gauge.
 Comm. Math. Phys. 82 (1982) 171-212
\bibitem{[Hormander]}
 H\"ormander L:
 The Analysis of Linear Partial Differential Operators II.
 Differential Operators with Constant Coefficients.
 Grundl. d. math. Wiss. 256,
 Springer-Verlag 1971, Berlin-Heidelberg 1983
\bibitem{[IsBaYa]}
Isenberg J, Bao D, Yasskin P B:
 Classical Supergravity. In:
 Mathematical Aspects of Superspace (ed.: H. J. Seifert et al).
 Nato Asi Series C: Math. and Phys. Sciences.
 Dordrecht 1984, 173--205
\bibitem{[KostBRST]}
Kostant B, Sternberg S:
 Symplectic reduction, BRS cohomology, and infinite-dimensional
 Clifford algebras.
 Annals of Physics 176, 49--113 (1987)
\bibitem{[Reed-Simon]}
Reed M, Simon B:
 Methods of Modern Mathematical Physics.
 II. Fourier Analysis, Self-Adjointness.
 Academic Press, New York, 1975
\bibitem{[Segal]}
Segal I:
 Symplectic Structures and the Quantization Problem for Wave Equations.
 Symposia Math. 14 (1974), 99-117
\bibitem{[HERM]}
Schmitt T:
 Supergeometry and hermitian conjugation.
 Journal of Geometry and Physics, Vol. 7, n. 2, 1990
\bibitem{[CMP1]}
 \bysame:
 Functionals of classical fields in quantum field theory.
 Reviews in Mathematical Physics,  Vol. 7, No. 8 (1995), 1249-1301
\bibitem{[CMP2]}
 \bysame:
 Supergeometry and quantum field theory, or:
 What is a classical configuration?
 Preprint No. 419, Techn. Univ. Berlin, 1995
\bibitem{[IS]}
 \bysame:
 Infinitedimensional Supermanifolds I.
 Report 08/88 des Karl-Weierstra\ss-Instituts f\"ur Mathematik,
 Berlin 1988.
 \par II, III.
 Mathematica Gottingensis. Schriftenreihe des SFBs Geometrie und Analysis,
 Heft 33, 34 (1990). G\"ottingen 1990
\bibitem{[Sniatycki]}
\'Sniatycki J, Schwarz G:
 The existence and uniqueness of solutions of Yang-Mills equations with
 the bag boundary conditions.
 Comm. Math. Phys. 159 (1994), 593--604.
\bibitem{[Taylor]}
Taylor M E:
 Pseudodifferential Operators.
 Princeton Math. Series, Princeton University Press, Princeton 1981
\bibitem{[Hd2IsAlg]}
Strichartz R S:
 Multipliers on fractional Sobolev spaces.
 J. Math. Mechanics 16 (1967), 1031 -- 1060
\end{thebibliography}
\end{document}